\documentstyle[11pt]{article}
\begin{document}
\begin{titlepage}
\begin{flushright}
YITP-00-06\\
nucl-th/0004056\\
April 2000
\end{flushright}
\begin{centering}
 
{\ }\vspace{1cm}
 
{\Large\bf Nuclear Muon Capture on the Proton and $^3$He}\\
\vspace{0.5cm}
{\Large\bf within the Standard Model and Beyond}\\
\vspace{2.0cm}
Jan Govaerts$^{\dag ,}$\footnote{On leave from the Institute of Nuclear
Physics, Catholic University of Louvain, Louvain-la-Neuve, Belgium\\
\indent{\ }\hspace{5pt}E-mail: {\tt govaerts@fynu.ucl.ac.be}}
and Jose-Luis Lucio-Martinez$^{\star ,}$\footnote{On leave from the
Instituto de F\'{\i}sica, Universidad de Guanajuato,
P.O. Box E-143, 37150 Le\'on, M\'exico\\
\indent{\ }\hspace{5pt}E-mail: {\tt lucio@ifug3.ugto.mx}}\\
\vspace{0.6cm}
$^{\dag}$
{\em C.N. Yang Institute for Theoretical Physics}\\
{\em State University of New York at Stony Brook}\\
{\em Stony Brook NY 11794-3840, USA}\\
\vspace{0.5cm}
$^{\star}$
{\em INFN-Laboratori Nazionale di Frascati}\\
{\em P.O. Box 13}\\
{\em I-00044 Frascati (Roma), Italy}
\vspace{2cm}
\begin{abstract}

\noindent Nuclear muon capture on the proton and $^3$He is considered
both within and beyond the Standard Model in terms as general as is
possible. Explicit and precise analytic expressions for all possible 
observables are given,
assuming only a Dirac neutrino in the limit of vanishing mass.
These results allow both for precision tests of the Standard Model and
new physics, as well as for the assessment of the potential physics reach
of experiments designed to measure specific observables.
Using these expressions, stringent constraints can already be inferred
from a recent precision measurement of the statistical capture rate
on $^3$He. Likewise, similar constraints should follow the completion
of a precision measurement in progress of the singlet capture rate
on the proton.

\end{abstract}

\vspace{10pt}

\end{centering} 

\vspace{5pt}


\vspace{25pt}

\end{titlepage}

\setcounter{footnote}{0}

\section{Introduction}
\label{Sect1}

Over the last fifteen years, muon physics has regained its rightful place 
in particle physics, ranging from intermediate energies well into the
high energy frontier in the foreseeable future with the advent of
muon colliders. Given the availability of intense muon beams at
different laboratories, as well as new and much efficient experimental and
detector techniques, intermediate energy muon physics has moved into
the realm of precision studies of the Standard Model, with the hope of
possibly unravelling some tell-tale sign for the physics which must
lie beyond it in ways complementary to present day high energy experiments
at the colliders. Such studies include both purely leptonic as well as
semi-leptonic electroweak processes, the much studied field of nuclear 
muon capture\cite{Mukho} belonging to the latter class.

When the aim of such electroweak processes in nuclei
is to study particle physics issues, the uncertainties
inherent to nuclear structure modeling have to be disposed of
to the largest extent possible, leaving essentially only the lightest
of nuclei available, beginning with the proton and next the stable
3-nucleon bound state, $^3$He, since indeed the quantum wave functions
for the latter system are by now very well understood\cite{JimPhD,Congleton}. 
An additional bonus in the case of these two systems is that they both
provide the unique occurrence of a (mother-daughter) pair of nuclei
which define a spin 1/2 isospin doublet, namely the proton and neutron,
and $^3$He and $^3$H, respectively. Hence in the limit of exact isospin 
symmetry, some of the various phenomenological nuclear form factors 
which parametrize matrix elements of these states 
are related to one another, while for spin 1/2 states, the numbers of these
form factors remains small. This is how through CVC, the experimental 
determination of electromagnetic form factors from electron scattering
may be translated into knowledge of the related nuclear form factors for
the charged electroweak vector current. In other words, hiding our
ignorance of the microscopic dynamics at the quark level into a
phenomenological parametrization in terms of form factors, it remains
possible to consider predictions of observables which do not require
models for nuclear structure. In the field of nuclear muon capture,
this description corresponds to the so-called
``elementary particle model" approach to nuclear electroweak
processes\cite{Kim2}, which will be used in this paper.

One of the main motivations for studying nuclear muon capture, especially 
in light nuclei\footnote{In heavier nuclei, the same axial electroweak 
probe enables to address the issue of hadronic coupling and mass 
renormalization in the nuclear medium.}, has always been to measure the 
induced pseudoscalar nucleon form factor $g_P(q^2)$, certainly the least 
well-known of all non vanishing nucleon form factors with a combined 
uncertainty which has stood at 22\% for the last twenty years\cite{Bardin}. 
The urgency of this specific issue has recently become more pressing, mainly 
for two reasons. On the one hand, based on the fundamental chiral symmetries 
which, even though dynamically broken, survive non linearly in the 
nonperturbative low energy regime of the theory for the strong
interactions among quarks, namely quantum chromodynamics, definite predictions
for $g_P$ with a precision of a few percents or better have been
achieved\cite{Bernard1,Fearing,Bernard3,Bernard4}. The experimental 
confirmation of the expected value is thus a crucial low energy test 
of our basic concepts for the theory of the strong interactions.
On the other hand, a new 8\% precise measurement of $g_P$
has been completed in the intervening years, using the rare
process of radiative muon capture on hydrogen\cite{Jonkmans,Wright}.
The unsettling fact is that the value obtained differs significantly
(by a 4.2 $\sigma$ deviation) from the expected value. No theoretical 
explanation for this discrepancy having been found so far (see 
for example Ref.\cite{Bernard4} for references), the issue thus remains
open and this difficult experiment shrouded with some feeling of
uncertainty. Indeed, the general understanding of the dynamical breaking
of the chiral symmetries of QCD has so far never been found lacking 
in essentially any other low energy process.

During the same period, another precision measurement
of nuclear muon capture on $^3$He has also been completed\cite{Ackerbauer}.
In spite of its remarkable precision of 0.4\% for the statistical capture 
rate at $\lambda^{\rm stat}_{\rm exp}=1496\pm 4$ s$^{-1}$,
whose value is found to be in perfect agreement with the theoretical 
prediction 
$\lambda^{\rm stat}_{\rm theor}=1496\pm 21$ s$^{-1}$\cite{JimPhD,Congleton}, 
the ensuing value for $g_P$, though at 
exactly the expected level, is still precise to only 19\%\cite{Congleton2}.
One should add however that in terms of the corresponding pseudoscalar
{\sl nuclear\/} form factor $F_P(q^2)$ for $^3$He, the same experimental
result translates\cite{Ackerbauer,Vorobyov,Gov1,Gov2,Gov3}
into a 13\% precise test of the prediction based
on PCAC, the symmetry which together with CVC has historically
been the precusor for the chiral symmetries of QCD. Even though these 
results are still
a long way off the precision reached by the latest theoretical 
analyses\cite{Bernard1,Fearing,Bernard3,Bernard4}, they stand as a clear
confirmation of the basic concepts involved in the chiral symmetry aspects
of the problem.

Independently of this specific situation with respect to the value for
$g_P$, if one is willing to use the theoretically expected number, the latest
ordinary muon capture experiment on $^3$He has reached such a level of
precision that other tests of the Standard Model (SM) for the electroweak
interactions become possible\cite{Gov1,Gov2,Gov3}, some of which prove
to be quite stringent for possible new physics beyond the SM.
Moreover, given the above issues surrounding the value of
$g_P$, as well as the precision of its theoretical prediction, a new effort
has been launched\cite{muH,Vorobyov} in order to measure the singlet rate of
ordinary muon capture on hydrogen in a gas target (to avoid the complications 
due to molecular binding effects), hopefully to a precision of 0.5\%
to 1\%\footnote{The principle of the experiment lies in the comparison
between the muon disappearance rates, through the usual electron decay mode,
for both positive and negative muons, only the latter being subject
to nuclear capture. Since this experiment does not measure the neutron 
recoil energy distribution, it has no handle on the neutrino energy spectrum 
nor on its mass.}.  Here again, beyond the initial aim towards the value 
of $g_P$, this level of precision should also enable tests of the SM.
However, in order to infer a value for $g_P$ from any experiment,
analytic expressions for observables whose numerical evaluation is to
the required standard of precision should be available. This is the
purpose of the present work, much in continuation of that of 
Refs.\cite{JimPhD,Congleton}.

This paper considers the ordinary nuclear muon capture process on a spin 1/2
isospin doublet in terms as general as is possible. Not only are all
possible effects existing within the SM included in the present
analysis, but any possible contributions which may appear beyond the SM
due to new interactions are accounted for as well through an effective
four-fermi quark-lepton interaction which describes all possible scalar,
pseudoscalar, vector, axial and tensor couplings. Indeed, given the
momentum transfer involved, much less than any of the mass scales
characterizing such possible interactions, an effective four-fermi
interaction at the quark-lepton level is perfectly justified.

Previous analyses over the years, beginning with Ref.\cite{Bernstein} pointing
out the existence of the hyperfine effect in the capture rates because
of maximal $(V-A)$ parity violation\footnote{The same effect arises of course
also for a pure $(V+A)$ coupling.}, have all only included vector and axial 
interactions through the usual $(V-A)$ charged electroweak coupling,
and only some of these works have considered the possibility of 
so-called second-class currents. A handful of observables have been computed,
and then not always in analytic form\cite{Kim2,Holstein1,
Santisteban,Hwang,Holstein3,JimPhD,Congleton,Congleton2,Congleton3,Bernard4}.
In contradistinction, the present work does not involve any non relativistic
expansion and allows for all possible contributions,
including $CP$- or $T$-odd effects through complex form factors
and couplings coefficients in the effective interaction. The only 
approximations made are, on the one hand, that the neutrino mass is taken 
to vanish and that the muon leptonic flavour is conserved, and on the other
hand, that nuclear recoil effects associated to the scalar, pseudoscalar
and tensor interactions beyond the SM are also ignored, 
since the latter couplings are necessarily small on their own and would 
in turn multiply small nuclear recoil contributions. 
Moreover, all possible observables
become available through our results, hopefully making it easier
to assess their sensitivity to whatever parameter or physical input
a specific experiment using ordinary nuclear muon capture on these nuclei would
wish to address. This point will explicitly be illustrated with some
of the obtained observables, in the context of the two experiments
mentioned previously.

Most of these observables require correlation measurements using
polarized states for the initial muonic atom, while another subset 
requires to measure also the
polarization states of the outgoing nucleus and/or neutrino. These are
certainly experimental challenges bordering on the impossible for certain
of these polarization observables, but some experimentalists take
up the task. For example, there exists a first-generation 
experiment\cite{Souder3} which measures the vector analyzing power of the
outgoing triton in polarized muon capture on $^3$He. Even though the
preliminary results are yet not precise enough to be of use in a theoretical
analysis, they certainly demonstrate that this specific difficult
challenge can be met.

More specifically, the physics reason for these difficulties is that during 
its atomic cascade down to the muonic atom ground state, the muon, even if 
initially polarized, suffers depolarization effects to a great extent, leaving 
over only a small fraction of its initial 
polarization\cite{Mukho,Favart,Souder2,Souder1}.
Once in the ground state, the degree of polarization may be increased
again by external means\cite{Kadano,Souder3}, but not to any large degree
and less than theoretically anticipated\cite{Kuno}. In fact, both the 
initial nucleus and muon need to be polarized in order to end up with a 
muonic atom ground state polarized to any degree\cite{Congleton1}.
Hence, because of these atomic physics issues, even though the expressions
for all observables are now available in analytic form, any experiment 
which consists not only in a rate measurement is in essence extremely 
difficult to perform when aiming towards the demands of great 
precision\footnote{In this respect, the recent proposal of Ref.\cite{Sobkow} 
appears to be totally unrealistic, the more so since some of the possible 
contributions which have not been included in that analysis, such as recoil 
order effects, could also lead to contributions to the massive neutrino 
polarization states. Such effects are all accounted for in our analysis
in the limit of a massless neutrino, including the possibility of 
$T$-violating couplings and currents other than $V$ and $A$. Note also 
that due to helicity constraints, transverse contributions to a massive 
neutrino polarization state produced in muon capture through only $V$ and $A$ 
couplings are necessarily suppressed by the ratio of the neutrino mass 
to its energy, and are thus unobservable.}.

The outline of the paper is as follows. In the next section, the general
parametrization of the capture amplitude is described, and the ensuing
expressions for observables explicitly given. In Section 3, the
situation for $^3$He is then considered in detail, first within the SM,
and then beyond the SM, each time by using the experimental result
of Ref.\cite{Ackerbauer} for the statistical capture rate and also by
illustrating the potential physics reach of some other final state distribution
which does not entail a correlation observable except for an initially
polarized muonic atom. In Section 4, the same considerations are applied
again to the case of capture on the proton, based on the optimistic
aim of a 0.5\% precision in the result for the singlet capture 
rate\cite{muH,Vorobyov}. Section 6 then translates some of the limits for 
physics beyond the SM obtained from the previous considerations, into limits 
for parameters of some specific models for such physics. The Conclusions end 
our discussion, while further information relevant to the analysis is 
provided in three separate Appendices.

\section{Muon Capture Observables}
\label{Sect2}

\subsection{Kinematics}
\label{Sect2.1}

Our notations for kinematics are as follows. Let $m_\mu$, $M_1$
and $M_2$ be the masses of the muon, and of the initial and final nuclei,
respectively, with $M=(M_1+M_2)/2$ the mean value of the latter two. 
As mentioned previously, the capture process is considered
in the limit of a vanishing neutrino mass and no leptonic flavour
mixing. Initially, the muon and nucleus form a muonic atomic bound state
at rest, whose total rest-mass is denoted $\sqrt{s}$, which differs
from $m_\mu+M_1$ by the binding energy $-(\alpha Z)^2\mu/2$, $\mu$ being
the reduced mass of the bound state (expressions are given in units such 
that $c=1$ and $\hbar=1$ throughout). Let then $\vec{p}$ be the 
momentum of the outgoing nucleus ($-\vec{p}$ is thus that of the outgoing 
neutrino), $\omega$ the energy of that nucleus and $\nu$ the energy of 
the neutrino, such that we have
\begin{equation}
\nu=|\vec{p}|\ \ ,\ \ 
\sqrt{s}=\nu+\omega\ \ ,\ \ 
\nu=\frac{s-M^2_2}{2\sqrt{s}}\ \ ,\ \ 
\omega=\frac{s+M^2_2}{2\sqrt{s}}. 
\end{equation}

With respect to polarization states, let $\hat{s}_1$ denote the normalized
polarization vector of the initial spin 1/2 nucleus, as measured in its
rest-frame, and $\hat{s}_2$ that of the final spin 1/2 nucleus {\sl also
measured in its rest-frame\/}. Similarly, $\hat{s}_\mu$ denotes the muon
normalized polarization vector in its rest-frame, while $\lambda=\pm 1$
is the massless Dirac\footnote{In practice, there is no difference 
between a massless Dirac or Majorana neutrino in the case of purely
$(V-A)$ or $(V+A)$ interactions, but the distinction becomes relevant
as soon as other interactions are turned on, as done in this
paper. By this assumption of a massless Dirac neutrino, we exclude the
possibility of interference contributions between processes in which
either the neutrino spinor field $\nu(x)$ or its charge conjugate $\nu^c(x)$
would couple in the amplitudes {\sl to the same quarks and leptons
rather than their antiparticles\/}, namely it is assumed that the muon
leptonic flavour is conserved in the massless neutrino limit.} 
neutrino helicity.

Finally, the capture distribution is given by the expression,
\begin{equation}
\frac{d\Gamma_\lambda}{d\Omega_{\hat{p}}}=\frac{|\psi_c(0)|^2C}
{64\pi^2m_\mu M_1}\frac{\nu}{\sqrt{s}}|{\cal M}_\lambda|^2.
\label{eq:kinerate}
\end{equation}
Here, $d\Omega_{\hat{p}}$ is the element of solid angle associated to the
outgoing nucleus of momentum $\vec{p}$,
${\cal M}_\lambda$ is the capture amplitude associated to a neutrino
of helicity $\lambda$, $\psi_c(0)$ is the 1S state muonic atom Coulomb
wave function measured at the origin, and $C$ is a reduction factor which 
accounts for the effects of the nuclei finite size through the overlap
of the muon and neutrino wave functions with the different nuclear
electric and electroweak charge distributions of finite spatial extent.
The detailed evaluation of this reduction factor $C$ is discussed
in Appendix 1 both for $^3$He and for the proton.

\subsection{The amplitude}
\label{Sect2.2}

The parametrization for the effective interaction associated
to muon capture, at the level of the $u$ and $d$ quarks, is
taken to be, 
\begin{displaymath}
4\frac{g^2}{8M^2}V_{ud}\sum_{\eta_1,\eta_2=+,-}\,\Bigg[\
\left({h^V_{\eta_1\eta_2}}\right)^*\bar{\nu}_\mu\gamma^\mu P_{\eta_1}\mu\,
\bar{d}\gamma_\mu P_{\eta_2}u\,+\,
\left({h^S_{\eta_1\eta_2}}\right)^*\bar{\nu}_\mu P_{\eta_1}\mu\,
\bar{d} P_{-\eta_2}u\,+\,
\end{displaymath}
\begin{equation}
+\frac{1}{2}
\left({h^T_{\eta_1\eta_2}}\right)^*\bar{\nu}_\mu\sigma^{\mu\nu}P_{\eta_1}\mu\,
\bar{d}\sigma_{\mu\nu} P_{-\eta_2}u\ \Bigg].
\label{eq:effec}
\end{equation}
Here, $P_\pm=(1\pm\gamma_5)/2$ are the chirality projectors, and $g$,
$M$ and $V_{ud}$ are arbitrary real parameters which in the limit of the
SM reduce to those of that Model, namely $M=M_W$, $g^2/8M^2=G_F/\sqrt{2}$
and $V_{ud}=\cos\theta_c$ being the Cabibbo-Kobayashi-Maskawa (CKM) 
quark flavour mixing matrix element, in which case all coefficients 
$h^{S,V,T}_{\pm\pm}$ also vanish except for $h^V_{--}=1$. Finally, the
coefficients $h^{V,S,T}_{\eta_1\eta_2}$ are arbitrary complex coefficients
associated to vector, scalar and tensor interactions, 
with $\eta_1$ (resp. $\eta_2$) being the muon 
(resp. $d$ quark) chirality, equal to the neutrino (resp. $u$ quark) one
for vector interactions and opposite to it for scalar and tensor interactions
(the ``$^*$" symbol denotes complex conjugation throughout).
Finally, without loss of generality, one may set $h^T_{++}=0=h^T_{--}$,
because of the identity
$\sigma^{\mu\nu}\gamma_5=i\epsilon^{\mu\nu\rho\sigma}\sigma_{\rho\sigma}/2$
which implies that the terms multiplied by $h^T_{++}$ and $h^T_{--}$
simply vanish identically.

This choice of parametrization is inspired by the one used (in the charge
exchange form) for muon decay in terms of coefficients 
$g^{S,V,T}_{\pm\pm}$, with the first (resp. second) lower
index being the chirality of the electron (resp. muon)\cite{PDG}.
Note that a similar effective four-fermi interaction may be given for
$\beta$-decay in terms of coefficients $f^{S,V,T}_{\pm\pm}$ with the 
electron then playing the role of the muon in (\ref{eq:effec}).

To express the amplitude for nuclear muon capture in terms of the above
parametrization, one also requires the hadronic matrix elements
of the relevant quark operators in terms of the nuclear
bound states. For this purpose in the case of a spin 1/2 isodoublet, let us 
introduce the spinor $\psi_1$ for the Dirac field of the
spin 1/2 nucleus on which the muon is captured, while $\psi_2$ is the
Dirac spinor for the final spin 1/2 nucleus. In momentum space,
with $q^\mu=p^\mu_2-p^\mu_1$ being the momentum transfer of the process
and $p^\mu_2$ (resp. $p^\mu_1$) being the momentum of the final
(resp. initial) nucleus, we have the following parametrization
in terms of $q^2$-dependent form factors,
\begin{equation}
<2|\overline{d}\gamma_\mu u|1>=\overline{\psi_2}
\left(F_V\gamma_\mu+iF_M\sigma_{\mu\nu}\frac{q^\nu}{2M}+
F_S\frac{q_\mu}{2M}\right)\psi_1,
\end{equation}
\begin{equation}
<2|\overline{d}\gamma_\mu\gamma_5 u|1>=\overline{\psi_2}
\left(F_A\gamma_\mu\gamma_5+F_P\gamma_5\frac{q_\mu}{m_\mu}+
iF_T\sigma_{\mu\nu}\gamma_5\frac{q^\nu}{2M}\right)\psi_1,
\end{equation}
\begin{equation}
<2|\overline{d} u|1>=\overline{\psi_2}\left(G_S\right)\psi_1,
\end{equation}
\begin{equation}
<2|\overline{d}\gamma_5 u|1>=\overline{\psi_2}\left(G_P\gamma_5\right)
\psi_1,
\end{equation}
\begin{equation}
<2|\overline{d}\sigma_{\mu\nu} u|1>=\overline{\psi_2}
\left(G_T\sigma_{\mu\nu}\right)\psi_1\ \ ,\ \ 
<2|\overline{d}\sigma_{\mu\nu}\gamma_5 u|1>=\overline{\psi_2}
\left(G_T\sigma_{\mu\nu}\gamma_5\right)\psi_1, 
\end{equation}
where $|1>$ and $|2>$ denote the initial and final quantum nuclear states,
respectively.

This notation for form factors is the one used for the $^3$He-$^3$H case,
while it is more conventional in the proton-neutron case to denote
the form factors for the vector and axial quark currents as
$g_V$, $g_M$, $g_S$, $g_A$, $g_P$ and $g_T$, respectively for
$F_V$, $F_M$, $F_S$, $F_A$, $F_P$ and $F_T$. We shall thus follow that
convention in the case of the proton, and keep nevertheless the notation
$G_S$, $G_P$ and $G_T$ for the form factors associated to the
scalar, pseudoscalar and tensor quark operators even in the case of the
proton. However, all expressions for observables in this section will
be given in terms of the $F_{V,M,S,A,P,T}$ form factors. 
Note also that the contribution
of the induced pseudoscalar form factor $F_P$ has conventionally been
normalized to the muon mass, while those of all other recoil order
form factors are normalized with respect to twice the mean nuclear mass $2M$.

In the case of $T$-invariant interactions, both the effective
coupling coefficients $h^{S,V,T}_{\pm\pm}$ as well as the nuclear
form factors are all real quantities under complex conjugation.
However in the present analysis, this restriction is not imposed,
and all these parameters are assumed to take {\sl a priori\/} complex values.

For the vector and axial current quark operators, one distinguishes
first- and second-class form factors, $F_V$, $F_M$, $F_A$
and $F_P$ in the first case, and $F_S$, $F_T$ in the second case.
On the basis of CVC, the values for the electroweak vector and
induced magnetic form factors may be related to those of the
electromagnetic electric and magnetic form factors of $^3$He and $3$H,
or of the proton and the neutron. The value of the axial form factor
$F_A$ at zero momentum transfer follows from the $\beta$-decay rate
either of $^3$H or of the neutron. Its value at non vanishing momenta
transfers requires knowledge of its $q^2$-dependency inferred from
neutrino or pion electroproduction experiments. Finally, through PCAC, the
value of the last first-class form factor, the induced pseudoscalar one,
may be expressed as\footnote{To be precise, these expressions for $F_P$
assume implicitly that both $q^2$-dependencies for $F_A(q^2)$ and
$g_{\pi N}(q^2)$ are identical\cite{Klieb}. Note also that through
relations such as these, any experiment leading to a precise value
for $F_P$ would also imply a precise value for the associated
pion-nucleus coupling constant\cite{Mukho1}.}
\begin{equation}
F^{\rm PCAC}_P(q^2)=\frac{m_\mu(M_1+M_2)F_A(q^2)}{m^2_\pi-q^2}=
\frac{m_\mu f_\pi g_{\pi N}(q^2)}{m^2_\pi-q^2},
\label{eq:PCAC}
\end{equation}
where $m_\pi$ is the $\pi^\pm$ mass, $f_\pi$ its decay constant, and
$g_{\pi N}(q^2)$ its nuclear coupling to the two nuclear states of masses
$M_1$ and $M_2$.
In the limit of exact isospin symmetry, which implies also exact
$G$-parity invariance, the values for the second-class form factors
$F_S$ and $F_T$ may be shown to vanish identically. Hence, one expects
deviations from zero for these form factors (normalized to $F_V$ or $F_A$)
of only a few percent, as given by the ratio $(m_d-m_u)/\Lambda_{\rm QCD}$ 
of $u$ and $d$ quark masses to the QCD scale for example, or the ratio
$(m_n-m_p)/(m_n+m_p)$ in terms of the neutron and proton masses. 
Bag model or QCD sum rule evaluations of these
form factors in the case of the (proton,neutron) doublet do indeed
bear out such an expectation\cite{Donoghue,Holstein2,Dominguez,Shiomi}.
In the same manner, one could wonder about the effects of isospin
symmetry breaking for the $F_V$ and $F_M$ form factors obtained
through CVC. In this particular case, thanks to the Ademollo-Gatto
theorem\cite{AG}, isospin breaking corrections are only of quadratic order 
in the ratio $(m_d-m_u)/\Lambda_{\rm QCD}$, hence negligeable for 
our purposes. Let us also recall that CVC has been very well established
through precision studies in $\beta$-decay\cite{CVC}, and that
even though limits exist on second-class form factors from correlation
experiments in $\beta$-decay, the stringency of these
constraints are nowhere close to the theoretically expected 
values for $g_S$ or $g_T$\cite{PDG}.

In the matrix elements for the scalar, pseudoscalar and tensor quark 
operators, we have chosen not to include
recoil order induced contributions, for the reasons mentioned
already in the Introduction. The values for the genuine
scalar, pseudoscalar and tensor form factors $G_S$, $G_P$ and $G_T$,
cannot be inferred from any experiment yet. One thus has to rely on
specific model calculations for QCD dynamics, such as the bag model
or QCD sum rules. However, no such results are available at present,
and one may only reasonably guess that these form factors should
take values on the order of unity, within a factor which at worst
could be of order ten.

Hence, even though the parametrization of the nuclear matrix elements
of the relevant quark operators in terms of form factors is only a
phenomenological representation of our ignorance of the microscopic
nonperturbative quark dynamics, this approach to nuclear muon capture allows 
nevertheless for explicit predictions independently of the details of nuclear 
mo\-dels, relying only on the results of other experiments and the power of
symmetry principles\cite{Kim2}.

In terms of this parametrization of the nuclear state matrix elements,
the effective muon capture amplitude at the nuclear level is given by,
\begin{displaymath}
{\cal M}_\lambda=\frac{g^2}{8M^2}V_{ud}\sum_{\eta_1,\eta_2=+,-}\,\Bigg[\
\left({h^V_{\eta_1\eta_2}}\right)^*\bar{\nu}_\mu\gamma^\mu
(1+\eta_1\gamma_5)\mu\,\times
\end{displaymath}
\begin{displaymath}
\times\,\overline{\psi_2}
[\gamma_\mu(F_V+\eta_2F_A\gamma_5)+
\frac{q_\mu}{2M}(F_S+\eta_2\frac{2M}{m_\mu}F_P\gamma_5)+
i\sigma_{\mu\nu}\frac{q^\nu}{2M}(F_M+\eta_2F_T\gamma_5)]\psi_1\,+\,
\end{displaymath}
\begin{displaymath}
+\,\left({h^S_{\eta_1\eta_2}}\right)^*\bar{\nu}_\mu(1+\eta_1\gamma_5)\mu\ 
\overline{\psi_2}(G_S-\eta_2G_P\gamma_5)\psi_1\,+\,
\end{displaymath}
\begin{equation}
+\,\frac{1}{2}\left({h^T_{\eta_1\eta_2}}\right)^*
\bar{\nu}_\mu\sigma^{\mu\nu}(1+\eta_1\gamma_5)\mu\ 
\overline{\psi_2}G_T\sigma_{\mu\nu}(1-\eta_2\gamma_5)\psi_1\,\Bigg].
\label{eq:M}
\end{equation}

\subsection{The method of calculation}
\label{Sect2.3}

The remainder of the calculation requires now the evaluation of
$|{\cal M}_\lambda|^2$. If one were to proceed by ``brute force", using the
usual trace techniques for such calculations, one would quickly run
into unmanageable expressions, because of the many contributions stemming
from all the interactions represented in the amplitude ${\cal M}_\lambda$.
Actually, it is possible in the present case to take advantage of the fact
that the initial system is at rest, and that to a very good approximation
the initial muon and nucleus may be considered to be also 
at rest\footnote{This assumption amounts to ignoring those small relativistic
corrections which are related to the velocities of the bound muon and nucleus,
which in the present case is indeed totally justified.
A fully satisfactory relativistic treatment of such a bound state problem
in QED is still not available, and cannot be used here to estimate small
corrections which in any case will be at most of order $(\alpha Z)^2$,
namely the squared velocity of the bound muon.}.

For that purpose, it turns out that the Dirac representation of the
Dirac-Clifford algebra of $\gamma^\mu$ matrices is the best suited for
the problem. Plane wave solutions to the massive Dirac equation are then
of the following form, for ``positive" and ``negative" energy solutions, 
respectively (our Minkowski metric signature convention is $(+---)$),
\begin{equation}
u(\vec{k},\hat{s})=\frac{1}{\sqrt{k^0+m}}(k^\mu\gamma_\mu+m)
\left(\begin{array}{c}
	\chi_+(\hat{s})\\
	0
	\end{array}\right)\ \ ,\ \
v(\vec{k},\hat{s})=\frac{1}{\sqrt{k^0+m}}(-k^\mu\gamma_\mu+m)
\left(\begin{array}{c}
	0\\
	\chi_-(\hat{s})
	\end{array}\right)\ \ ,\ \
\end{equation}
where $\chi_\pm(\hat{s})$ are bi-spinors such that
\begin{equation}
\vec{\sigma}\cdot\hat{s}\,\chi_\pm(\hat{s})=\pm\,\chi_\pm(\hat{s})\ \ ,\ \ 
\chi^\dagger_\pm(\hat{s})\chi_\pm(\hat{s})=1,
\end{equation}
with $\sigma^i$ $(i=1,2,3)$ the usual Pauli matrices,
while $\hat{s}$ is a unit vector in three dimensions which in fact corresponds
to the spin polarization vector of the particle (see Appendix 2). 
Further properties of these bi-spinors are
\begin{equation}
\chi_+(-\hat{s})=i\chi_-(\hat{s})\ \ ,\ \ 
\chi_-(-\hat{s})=i\chi_+(\hat{s}), 
\end{equation}
\begin{equation}
\chi_\eta(\hat{s})\chi^\dagger_\eta(\hat{s})=
\frac{1}{2}\left[1+\eta\vec{\sigma}\cdot\hat{s}\right]\ \ ,\ \ \eta=\pm.
\label{eq:trace}
\end{equation}
In fact, it is only the latter relation which is of essential use in the
calculation of $|{\cal M}_\lambda|^2$.

Even though this is not important for that calculation, it is also
possible to give the explicit expressions for these bi-spinors.
Associated to the spherical angular parametrization of the unit vector 
$\hat{s}$,
\begin{equation}
\hat{s}\ \ :\ \ 
\left(\begin{array}{c}
	\sin\theta\cos\varphi\\
	\sin\theta\sin\varphi\\
	\cos\theta
	\end{array}\right),
\label{eq:spherical}
\end{equation}
one has
\begin{equation}
\chi_+(\hat{s})=\left(\begin{array}{c}
		e^{-i\varphi/2}\cos\theta/2\\
		e^{i\varphi/2}\sin\theta/2
			\end{array}\right)\ \ ,\ \ 
\chi_-(\hat{s})=\left(\begin{array}{c}
		-e^{-i\varphi/2}\sin\theta/2\\
		e^{i\varphi/2}\cos\theta/2
			\end{array}\right). 
\label{eq:chifunctions}
\end{equation}

Finally in the case of solutions to the massless Dirac equation,
``positive" and ``negative" energy plane wave solutions of helicity 
$\lambda=\pm 1$ and momentum $\vec{k}$ are given by
\begin{equation}
u(\vec{k},\lambda)=\sqrt{k^0}\left(\begin{array}{c}
			\chi_\lambda(\hat{k})\\
			\lambda\chi_\lambda(\hat{k})
				\end{array}\right)\ \ ,\ \ 
v(\vec{k},\lambda)=\sqrt{k^0}\left(\begin{array}{c}
			\chi_\lambda(\hat{k})\\
			\lambda\chi_\lambda(\hat{k})
				\end{array}\right), 
\end{equation}
where of course $\hat{k}=\vec{k}/|\vec{k}|$.

In order to apply these considerations to the capture amplitude
(\ref{eq:M}), let us denote by $\chi_\mu(\hat{s}_\mu)$, 
$\chi_1(\hat{s}_1)$, $\chi_2(\hat{s}_2)$ and 
$\chi_\lambda(-\hat{p})$ the bi-spinors associated to the muon, 
the initial and final nuclei, and the neutrino, respectively
(for the first three states, they thus correspond to bi-spinors of
$\chi_+(\hat{s})$ type).
After substitution in (\ref{eq:M}), the capture amplitude associated 
to a massles neutrino of helicity $\lambda$ then reduces to the expression,
\begin{displaymath}
{\cal M}_\lambda=\left(\frac{g^2}{8M^2}\right)\,V_{ud}\,\sqrt{4mM_1}
\sqrt{\frac{\nu}{2\sqrt{s}}}\times
\end{displaymath}
\begin{displaymath}
\times\,2\Bigg\{\chi^\dagger_\lambda\,\chi_\mu\
\chi^\dagger_2\Big[H^{(S)}_\lambda+(H^{(P)}_\lambda-H^{(A)}_\lambda)
\hat{p}\cdot\vec{\sigma}\Big]\,\chi_1\,+\,
\end{displaymath}
\begin{equation}
+\lambda\,\,\chi^\dagger_\lambda\,\sigma^i\,\chi_\mu\
\chi^\dagger_2\Bigg[H^{(V)}_\lambda\,i(\hat{p}\times\vec{\sigma})^i\,-\,
H^{(A)}_\lambda\,\sigma^i\Big]\chi_1\,\Bigg\},
\end{equation}
with coefficients $H^{(S,P,V,A)}_\lambda$ defined by the following relations
\begin{equation}
H^{(S)}_\lambda=
\left(\sum_{\eta_2=\pm}h^V_{\lambda\eta_2}\right)^*\,G^{(1)}_V\,+\,
\left(\sum_{\eta_2=\pm}h^S_{-\lambda\eta_2}\right)^*\,G^{(1)}_S\,-\,
2{(h^T_{-\lambda\lambda})}^*\,G^{(1)}_T,
\end{equation}
\begin{equation}
H^{(P)}_\lambda=
\left(\sum_{\eta_2=\pm}\eta_2\,h^V_{\lambda\eta_2}\right)^*\,G^{(1)}_A\,+\,
\left(\sum_{\eta_2=\pm}\eta_2\,h^S_{-\lambda\eta_2}\right)^*\,G^{(1)}_P\,-\,
2\lambda{(h^T_{-\lambda\lambda})}^*\,G^{(2)}_T,
\end{equation}
\begin{equation}
H^{(V)}_\lambda=
\left(\sum_{\eta_2=\pm}h^V_{\lambda\eta_2}\right)^*\,G^{(2)}_V\,-\,
2{(h^T_{-\lambda\lambda})}^*\,G^{(1)}_T,
\end{equation}
\begin{equation}
H^{(A)}_\lambda=
\left(\sum_{\eta_2=\pm}\eta_2\,h^V_{\lambda\eta_2}\right)^*\,G^{(2)}_A\,-\,
2\lambda{(h^T_{-\lambda\lambda})}^*\,G^{(2)}_T,
\end{equation}
in which the following combinations of form factors are introduced,
\begin{equation}
G^{(1)}_V=(\sqrt{s}-M_2)\left[F_V-\frac{\sqrt{s}-M_1}{2M}F_M\right]\,+\,
(\sqrt{s}+M_2)\left[F_V+\frac{\sqrt{s}-M_1}{2M}F_S\right],
\end{equation}
\begin{equation}
G^{(1)}_A=(\sqrt{s}-M_2)\left[F_A-\frac{\sqrt{s}-M_1}{m_\mu}F_P\right]\,+\,
(\sqrt{s}+M_2)\left[F_A+\frac{\sqrt{s}-M_1}{2M}F_T\right],
\end{equation}
\begin{equation}
G^{(2)}_V=(\sqrt{s}-M_2)\left[F_V+\frac{M_1+M_2}{2M}F_M\right],
\end{equation}
\begin{equation}
G^{(2)}_A=(\sqrt{s}+M_2)\left[F_A-\frac{M_1-M_2}{2M}F_T\right],
\end{equation}
\begin{equation}
G^{(1)}_S=(\sqrt{s}+M_2)G_S,
\end{equation}
\begin{equation}
G^{(1)}_P=(\sqrt{s}-M_2)G_P,
\end{equation}
\begin{equation}
G^{(1)}_T=(\sqrt{s}-M_2)G_T,
\end{equation}
\begin{equation}
G^{(2)}_T=(\sqrt{s}+M_2)G_T.
\end{equation}

The calculation of $|{\cal M}_\lambda|^2$ then proceeds using the
trace properties in (\ref{eq:trace}). The advantage of the above
approach is that the combinations of form factors and coupling coefficients
which are relevant appear from the start, ever before proceeding
to the calculation of traces. Were one to first calculate
the traces as is usual, the task would quickly become impossible in 
the general case considered here (presumably, this is the reason why only 
numerical expressions were given in Ref.\cite{Santisteban}, even though 
the situation considered there was far less general).
Still, using the present approach, the length of the calculation is of
some importance. Note also that no nonrelativistic expansion in the amplitude
is effected at any stage of the calculation, in contradistinction
to all other analyses which are based on Ref.\cite{Kim2} in which such
an expansion in $1/M$ is indeed applied.

Finally, the capture distribution is thus given by,
\begin{equation}
\frac{d\Gamma_\lambda}{d\Omega_{\hat{p}}}=
\frac{|\psi(0)|^2}{32\pi^2}\,\left(\frac{g^2}{8M^2}\right)^2\,
V^2_{ud}\,\frac{\nu^2}{s}\,R_\lambda,
\end{equation}
with,

\noindent $R_\lambda=$
\begin{displaymath}
\left(1-\lambda\hat{p}\cdot\hat{s}_\mu\right)\,
\Big\{C^{(\mu)}_\lambda\,
\left(1+(\hat{p}\cdot\hat{s}_1)(\hat{p}\cdot\hat{s}_2)\right)\,+\,
D^{(\mu)}_\lambda\,\left((\hat{s}_1\cdot\hat{s}_2)-
(\hat{p}\cdot\hat{s}_1)(\hat{p}\cdot\hat{s}_2)\right)\,+\,
E^{(\mu)}_\lambda\,\hat{p}\cdot(\hat{s}_1\times\hat{s}_2)\,\Bigg\} +
\end{displaymath}
\begin{displaymath}
+\ \left(1-\lambda\hat{p}\cdot\hat{s}_1\right)\,
\Big\{C^{(1)}_\lambda\,
\left(1+(\hat{p}\cdot\hat{s}_\mu)(\hat{p}\cdot\hat{s}_2)\right)\,+\,
D^{(1)}_\lambda\,\left((\hat{s}_\mu\cdot\hat{s}_2)-
(\hat{p}\cdot\hat{s}_\mu)(\hat{p}\cdot\hat{s}_2)\right)\,+\,
E^{(1)}_\lambda\,\hat{p}\cdot(\hat{s}_\mu\times\hat{s}_2)\,\Bigg\} +
\end{displaymath}
\begin{equation}
+\ \left(1+\lambda\hat{p}\cdot\hat{s}_2\right)\,
\Big\{C^{(2)}_\lambda\,
\left(1-(\hat{p}\cdot\hat{s}_\mu)(\hat{p}\cdot\hat{s}_1)\right)\,+\,
D^{(2)}_\lambda\,\left((\hat{s}_\mu\cdot\hat{s}_1)-
(\hat{p}\cdot\hat{s}_\mu)(\hat{p}\cdot\hat{s}_1)\right)\,+\,
E^{(2)}_\lambda\,\hat{p}\cdot(\hat{s}_\mu\times\hat{s}_1)\,\Bigg\},
\label{eq:distr}
\end{equation}
in which the following final definitions apply,
\begin{equation}
C^{(\mu)}_\lambda=
|H^{(S)}_\lambda|^2+|H^{(P)}_\lambda|^2-
2\,|H^{(V)}_\lambda+\lambda H^{(A)}_\lambda|^2,
\end{equation}
\begin{equation}
C^{(1)}_\lambda=
-2\lambda\, {\rm Re}\left(H^{(S)}_\lambda(H^{(P)}_\lambda)^*\right)+
2\,|H^{(V)}_\lambda+\lambda H^{(A)}_\lambda|^2,
\end{equation}
\begin{equation}
C^{(2)}_\lambda=
+2\lambda\, {\rm Re}\left(H^{(S)}_\lambda(H^{(P)}_\lambda)^*\right)+
2\,|H^{(V)}_\lambda+\lambda H^{(A)}_\lambda|^2,
\end{equation}
\begin{equation}
D^{(\mu)}_\lambda=|H^{(S)}_\lambda|^2-|H^{(P)}_\lambda|^2,
\end{equation}
\begin{equation}
D^{(1)}_\lambda=-2\,{\rm Re}\left((H^{(S)}_\lambda-\lambda H^{(P)}_\lambda)
(H^{(V)}_\lambda+\lambda H^{(A)}_\lambda)^*\right),
\end{equation}
\begin{equation}
D^{(2)}_\lambda=-2\,{\rm Re}\left((H^{(S)}_\lambda+\lambda H^{(P)}_\lambda)
(H^{(V)}_\lambda+\lambda H^{(A)}_\lambda)^*\right),
\end{equation}
\begin{equation}
E^{(\mu)}_\lambda=2\,{\rm Im}\left(H^{(S)}_\lambda(H^{(P)}_\lambda)^*\right),
\end{equation}
\begin{equation}
E^{(1)}_\lambda=-2\lambda\,{\rm Im}
\left((H^{(S)}_\lambda-\lambda H^{(P)}_\lambda)
(H^{(V)}_\lambda+\lambda H^{(A)}_\lambda)^*\right),
\end{equation}
\begin{equation}
E^{(2)}_\lambda=-2\lambda\,{\rm Im}
\left((H^{(S)}_\lambda+\lambda H^{(P)}_\lambda)
(H^{(V)}_\lambda+\lambda H^{(A)}_\lambda)^*\right).
\end{equation}

Note that these results show that $T$-odd effects can only
appear through the triple correlation coefficients of $E^{(\mu,1,2)}_\lambda$ 
type, related to contributions which are pure imaginary under complex 
conjugation and involving necessarily always at least two polarization vectors.

\subsection{Capture distributions and final state polarization}
\label{Sect2.4}

The calculation thus leads to the capture 
distribution as given in (\ref{eq:distr}). For a massless neutrino
of given helicity
$\lambda$ ({\em i.e.\/} a Weyl neutrino, or a neutrino whose helicity
is measured!), 
(\ref{eq:distr}) gives the final expression. For a massless
Dirac neutrino produced with either helicity (which is then not measured), 
one needs still to sum
the result over $\lambda=\pm 1$. Finally, and independently of the
neutrino sector, in order to obtain the unpolarised nuclear
final state distribution, one simply needs to sum the
two results obtained from (\ref{eq:distr})
for $\hat{s}_2=\hat{s}_0$ and $\hat{s}_2=-\hat{s}_0$. Since
the result (\ref{eq:distr})
is at most linear in $\hat{s}_2$, this amounts to setting
$\hat{s}_2=\vec{0}$ and to multiply the result (\ref{eq:distr})
by a factor two.

In general, (\ref{eq:distr}) thus has the parametrization,
\begin{equation}
R_\lambda={\cal N}_\lambda+\hat{s}_2\cdot\vec{\cal D}_\lambda,
\label{eq:ND}
\end{equation}
where the expressions for the coefficients ${\cal N}_\lambda$ and 
$\vec{\cal D}_\lambda$ may be read off directly from (\ref{eq:distr}).
The capture distribution of the final state nucleus is thus given by,
\begin{equation}
\frac{d\Gamma_\lambda}{d\Omega_{\hat{p}}}=\frac{|\psi(0)|^2}{32\pi^2}\,
\left(\frac{g^2}{8M^2}\right)^2\,V^2_{ud}\,\frac{\nu^2}{s}\,
2{\cal N}_\lambda,
\end{equation}
while the final state spin 1/2 nucleus has the following average
polarization vector,
\begin{equation}
<\hat{s}_2>_\lambda=\frac{1}{{\cal N}_\lambda}\,\vec{\cal D}_\lambda.
\label{eq:polari}
\end{equation}

For example taking an initially unpolarized state with 
$\hat{s}_\mu=\vec{0}=\hat{s}_1$, the average polarization
of the final nucleus is given by
\begin{equation}
<\hat{s}_2>^{({\rm unpolarized})}_\lambda=
\lambda\,\frac{C^{(2)}_\lambda}
{C^{(\mu)}_\lambda+C^{(1)}_\lambda+C^{(2)}_\lambda}\,\hat{p}.
\end{equation}
Hence, a transversally polarized final nucleus requires an initially
polarized muonic atom.

However, all the results discussed so far refer to polarization states
defined in terms of the individual polarization vectors $\hat{s}_\mu$,
$\hat{s}_1$ and $\hat{s}_2$, rather than the hyperfine states 
$(S,m)$ of the initial muonic atom, with $(S=1,m=\pm1,0)$ and $(S=0,m=0)$.
As explained in Appendix~2, it is of course possible to use the
above results to determine the final state distributions and polarizations
associated to each of these hyperfine states. Since it is difficult to 
imagine how
the final state polarization of the neutron or $^3$H could ever be measured
to any degree of precision, here only the relevant expressions for the
hyperfine capture distributions are presented.

Given the hyperfine states $S=0,1$ and their projections $m=0$ and
$m=0,\pm 1$ along some arbitrary quantization axis, the associated
hyperfine capture distributions are given by,
\begin{equation}
\frac{d\Gamma^{(S,m)}_\lambda}{d\Omega_{\hat{p}}}=
\frac{|\psi(0)|^2}{16\pi^2}\left(\frac{g^2}{8M^2}\right)^2V^2_{ud}
\frac{\nu^2}{s}\,R^{(S,m)}_\lambda,
\label{eq:hyper}
\end{equation}
with
\begin{equation}
R^{(1,\pm 1)}_\lambda=
\Big[(C^{(\mu)}_\lambda+C^{(1)}_\lambda)+\frac{2}{3}(C^{(2)}_\lambda
+D^{(2)}_\lambda)\Big]\,\mp
\,\lambda(C^{(\mu)}_\lambda+C^{(1)}_\lambda)P_1(\cos\theta)\,-
\,\frac{2}{3}(C^{(2)}_\lambda+D^{(2)}_\lambda)P_2(\cos\theta),
\end{equation}
\begin{equation}
R^{(1,0)}_\lambda=
\Big[(C^{(\mu)}_\lambda+C^{(1)}_\lambda)+\frac{2}{3}(C^{(2)}_\lambda
+D^{(2)}_\lambda)\Big]\,+
\,\frac{4}{3}(C^{(2)}_\lambda+D^{(2)}_\lambda)P_2(\cos\theta),
\end{equation}
\begin{equation}
R^{(0,0)}_\lambda=
(C^{(\mu)}_\lambda+C^{(1)}_\lambda)+2(C^{(2)}_\lambda
+D^{(2)}_\lambda)-4D^{(2)}_\lambda,
\end{equation}
where the usual Legendre polynomials are
\begin{equation}
P_1(\cos\theta)=\cos\theta\ \ \ ,\ \ \ 
P_2(\cos\theta)=\frac{3}{2}\cos^2\theta-\frac{1}{2},
\end{equation}
while $\theta$ is the angle between the spin quantization axis
and the normalized momentum vector $\hat{p}$
of the outgoing spin 1/2 nucleus.

In particular, the statistical capture distribution is given as in
(\ref{eq:hyper}) with the coefficient $R^{(S,m)}_\lambda$ then obtained from
\begin{displaymath}
R^{{\rm stat}}_\lambda=
\frac{1}{4}\Big[R^{(1,+1)}_\lambda+R^{(1,-1)}_\lambda+
R^{(1,0)}_\lambda+R^{(0,0)}_\lambda\Big]=
\end{displaymath}
\begin{equation}
=C^{(\mu)}_\lambda+C^{(1)}_\lambda+C^{(2)}_\lambda=
|H^{(S)}_\lambda|^2+|H^{(P)}_\lambda|^2
+2|H^{(V)}_\lambda+\lambda H^{(A)}_\lambda|^2,
\label{eq:Rstat}
\end{equation}
a results which is indeed $\theta$-independent as it should.

Integrating these distributions leads to the associated capture rates,
$\lambda^{\rm S}_\lambda$, $\lambda^{\rm T}_\lambda$ and
$\lambda^{\rm stat}_\lambda$ for the singlet (S), triplet (T) and
statistical rates, respectively, of the form
\begin{equation}
\lambda^{\rm S,T,stat}_\lambda=
\frac{|\psi(0)|^2}{4\pi}\left(\frac{g^2}{8M^2}\right)^2V^2_{ud}
\frac{\nu^2}{s}\,R^{\rm S,T,stat}_\lambda,
\label{eq:rates}
\end{equation}
with $R^{\rm stat}_\lambda$ for the statistical capture rate given
in (\ref{eq:Rstat}) already, while for the singlet and triplet capture
rates, one has,
respectively,
\begin{equation}
R^{\rm S}_\lambda=C^{(\mu)}_\lambda+C^{(1)}_\lambda+
2C^{(2)}_\lambda-2D^{(2)}_\lambda=R^{(0,0)}_\lambda,
\end{equation}
\begin{equation}
R^{\rm T}_\lambda=C^{(\mu)}_\lambda+C^{(1)}_\lambda+
\frac{2}{3}\left[C^{(2)}_\lambda+D^{(2)}_\lambda\right].
\end{equation}

It is also possible to represent the hyperfine capture distribution
(\ref{eq:hyper}) in the following form\cite{JimPhD,Congleton}
\begin{equation}
\frac{d\Gamma_\lambda}{d\Omega_{\hat{p}}}=\frac{1}{4\pi}
\lambda^{({\rm stat})}_\lambda\,
\left[1+A^\lambda_\Delta P_\Delta +A^\lambda_vP_v\cos\theta+
A^\lambda_t P_t(\frac{3}{2}\cos^2\theta-\frac{1}{2})\right],
\label{eq:hyper2}
\end{equation}
where $A^\lambda_\Delta$, $A^\lambda_v$ and $A^\lambda_t$ are specific 
coefficients, the latter
two known as the vector and tensor analyzing powers of the final state
nucleus, respectively, while $A^\lambda_\Delta$ is a measure of 
the hyperfine effect on the statistical capture rate since one has
\begin{equation}
A^\lambda_\Delta=\frac{1}{4}
\frac{\lambda^{\rm T}_\lambda-\lambda^{\rm S}_\lambda}
{\lambda^{\rm stat}_\lambda}=
\frac{\lambda^{\rm T}_\lambda-\lambda^{\rm S}_\lambda}
{3\lambda^{\rm T}_\lambda+\lambda^{\rm S}_\lambda}.
\end{equation}
Finally in (\ref{eq:hyper2}), the coefficients $P_{\Delta,v,t}$ 
are the following combinations of the hyperfine populations $N_{S,m}$
\begin{equation}
P_\Delta=N_{1,1}+N_{1,0}+N_{1,-1}-3N_{0,0}=1-4N_{0,0},
\end{equation}
\begin{equation}
P_v=N_{1,1}-N_{1,-1}\ \ ,\ \ 
P_t=N_{1,1}+N_{1,-1}-2N_{1,0},
\end{equation}
such that $N_{1,1}+N_{1,-1}+N_{1,0}+N_{0,0}=1$.

In terms of the quantities introduced above, one then finds,
\begin{equation}
A^\lambda_\Delta=-\frac{1}{3}\frac{C^{(2)}_\lambda-2D^{(2)}_\lambda}
{R^{\rm stat}_\lambda}\ \ ,\ \
A^\lambda_v=-\frac{\lambda(C^{(\mu)}_\lambda+C^{(1)}_\lambda)}
{R^{\rm stat}_\lambda}\ \ ,\ \
A^\lambda_t=-\frac{2}{3}\frac{C^{(2)}_\lambda+D^{(2)}_\lambda}
{R^{\rm stat}_\lambda}.
\end{equation}
Note that when observables are considered in which the summation
over the two neutrino helicity states $\lambda=\pm1$ has been effected,
this summation has to be applied separately in the numerator and the
denominator of each of the expressions given in this subsection.

\section{The $^3$He Case}
\label{Sect3}

\subsection{Physical inputs}
\label{Sect3.1}

The basic kinematical input used in the case of muon capture on $^3$He
is as follows,
\begin{equation}
\begin{array}{l c l c l}
M_1=2808.392\ {\rm MeV}\ \ &,&\ \ 
M_2=2808.928\ {\rm MeV}\ \ &,&\ \ 
\sqrt{s}=2914.039\ {\rm MeV},\\
\nu=103.22\ {\rm MeV}\ \ &,&\ \ \omega=2810.82\ {\rm MeV}\ \ &,&\ \ 
\omega-M_2=1.90\ {\rm MeV}.
\end{array}
\end{equation}

As explained in Appendix 1, the overlap reduction factor $C$ in this case
takes the value $C=0.979$\cite{JimPhD,Congleton}, while the latest
value quoted in Ref.\cite{PDG} for the CKM $ud$ mixing angle is used,
namely\footnote{The uncertainty on this value is irrelevant for the $^3$He
case, but must be included in the prediction for muon capture on the proton
at the precision level reached in Sect.\ref{Sect4.2}. Note also that
this value for $V_{ud}$ differs somewhat from that used 
previously\cite{Gov1,Gov2,Gov3}, which implies that the numbers quoted here
differ somewhat from those quoted earlier, but in no physically
significant way whatsoever.}
\begin{equation}
V_{ud}=0.9750\pm 0.0008.
\label{eq:Vud}
\end{equation}
This value results from a global fit which includes the constraints
of unitarity of the CKM mixing matrix, rather than the value
which would follow solely from the $0^+$-$0^+$ superallowed $\beta$-decay
$ft$-values.

Values for the nuclear form factors at the relevant momentum
transfer $q^2_1=-0.954m^2_\mu$ must also be specified. For this purpose,
we refer to the discussion in Refs.\cite{JimPhD,Congleton} and use the
values advocated by these authors. For the electroweak vector and induced
magnetic form factors, one has
\begin{equation}
F_V(q^2_1)=0.834\pm 0.011\ \ ,\ \ 
F_M(q^2_1)=-13.969\pm 0.052.
\end{equation}
The situation for $F_A(q^2_1)$ is somewhat more delicate, since only its
value at $q^2=0$ is known from $\beta$-decay of $^3$H, while its
$q^2$-dependency may only be inferred through a nuclear model calculation
of the associated mean square charge radius\cite{JimPhD,Congleton}.
This leads to the following value
\begin{equation}
F_A(q^2_1)=-1.052\pm[0.005-0.01],
\end{equation}
where the interval for the uncertainty is an attempt to reflect the lack
of precise knowledge of this form factor. The lower uncertainty of 0.005
stems from the uncertainties on the $^3$H $\beta$-decay rate, while the
corrections due to mesonic exchange currents may only be estimated in the
nuclear model calculation, leading to a total uncertainty of 0.007 on
$F_A(q^2_1)$\cite{JimPhD,Congleton}. Thus the bracket $[0.005-0.01]$
represents a conservative evaluation of the uncertainty on that form factor,
which will be carried along throughout our analysis later on.
As it turns out, the statistical capture rate is rather sensitive
to that quantity, so that any improvement on the evaluation of its
uncertainty leads to an improvement on the values of other quantities
that one infers from experiment. However, it is
difficult to see how such an improvement on $F_A(q^2_1)$ could be
achieved in practice, since it would require a measurement of the
$q^2$-dependency of the $^3$He-$^3$H axial form factor $F_A(q^2)$.

When it has to be specified, 
the value used for the induced pseudoscalar form factor
$F_P$ is that inferred from the PCAC relation (\ref{eq:PCAC}), which
corresponds to
\begin{equation}
F^{\rm PCAC}_P(q^2_1)=-20.73\pm[0.10-0.20],
\label{eq:FPPCAC}
\end{equation}
where the uncertainty range indicated in brackets thus corresponds to
the associated range in the value used for $F_A(q^2_1)$. 

Finally, the remaining two second-class form factors for the vector
and axial quark ope\-ra\-tors, $F_S(q^2_1)$ and $F_T(q^2_1)$, are taken to
be vanishing, even though for the reasons explained in Sect.\ref{Sect2.2},
one should expect that their absolute values,
normalized to those for $F_V(q^2_1)$ and $F_A(q^2_1)$, respectively, would 
be on the order of 0.02. As will become clear later on, the statistical
capture rate is in fact rather insensitive to these two form factors,
so that this reasonable approximation is certainly sufficient.

These are all the nuclear form factors required when considering
muon capture within the SM. Beyond that model however, the genuine
nuclear scalar, pseudoscalar and tensor form factors, $G_S(q^2_1)$,
$G_P(q^2_1)$ and $G_T(q^2_1)$, are also required. As mentioned previously,
values for these quantities are not known, but are expected to be on the
order of unity within a factor of at most ten. Nevertheless, any constraints 
put on the effective coupling coefficients $h^{S,V,T}_{\pm\pm}$ beyond the SM
will thus involve at present these nuclear form factors as well.

\subsection{Within the Standard Model}
\label{Sect3.2}

Capture within the SM corresponds to the amplitude ${\cal M}_\lambda$
in (\ref{eq:M}) with only the $h^V_{--}=1$ effective coupling coefficient
turned on. Given the above numerical values, as well as the general
expressions of Sect.\ref{Sect2.4}, it is straightforward to
determine theoretical values for all the quantities which enter
the general final state triton distribution associated to hyperfine
states as parametrized in (\ref{eq:hyper2}). One finds\footnote{Note that
the helicity index $\lambda$ is suppressed, since whether one sums or not
over the helicities of the final neutrino which is only of left-handed
chirality is irrelevant in the massless limit in the present case.},
\begin{equation}
\begin{array}{r c l}
\lambda^{\rm S}&=&1929\pm[28-46]\ {\rm s}^{-1},\\
\lambda^{\rm T}&=&1351\pm[16-19]\ {\rm s}^{-1},\\
\lambda^{\rm stat}&=&1496\pm[12-21]\ {\rm s}^{-1},\\
A_\Delta&=&-0.0967\pm[0.0061-0.0069],\\
A_v&=&+0.524\pm[0.0057-0.0061],\\
A_t&=&-0.3790\pm[0.00074-0.00123],
\end{array}
\end{equation}
where each time the indicated uncertainty includes all uncertainties
of all the theoretical input, while the range indicated by the bracket
corresponds to the range implied by the uncertainty on $F_A(q^2_1)$.
Obviously, some of these results coincide with the specific quantities
also computed in Refs.\cite{JimPhD,Congleton} in this particular case.

The disparity in the uncertainty ranges for these quantities stems from
their sensitivity to all theoretical inputs. This sensitivity of an
observable ${\cal O}$ to a parameter $F_X$ may be specified in terms of 
the variations, evaluated in the SM,
\begin{equation}
\sigma({\cal O};F_X)=\frac{F_X}{{\cal O}}
\frac{\partial{\cal O}}{\partial F_X}_{|_{\rm SM}}\ \ 
{\rm if}\ F_X\ne 0\ \ ;\ \ 
\sigma({\cal O};F_X)=\frac{1}{{\cal O}}
\frac{\partial{\cal O}}{\partial F_X}_{|_{\rm SM}}\ \ 
{\rm if}\ F_X=0\ \ . 
\label{eq:Sen}
\end{equation}
The values of these quantities for all the above hyperfine observables
are shown in Table \ref{Table1}. Note that except for $A_v$, all
these observables are rather sensitive to the axial form factor $F_A(q^2_1)$,
and thus to its bracket of uncertainty. Sensitivities to $F_P(q^2_1)$
include the results of Refs.\cite{JimPhD,Congleton}, showing
in fact that except for the triplet capture rate,
the statistical rate which has been measured\cite{Ackerbauer}
is the least sensitive observable
to that form factor, with the obvious drawback that all other observables
are still far more difficult to measure.

Given the experimental result for the statistical capture 
rate\cite{Ackerbauer}, 
\begin{equation}
\lambda^{\rm stat}_{\rm exp}=1496\pm 4\ {\rm s}^{-1},
\label{eq:exp3He}
\end{equation}
it is possible to infer a value for any given
parameter once the values for all other quantities are specified.
Thus given the inputs discussed in Sect.\ref{Sect3.1} except for the
value for $F_P(q^2_1)$, the result (\ref{eq:exp3He}) 
implies\cite{Ackerbauer, Vorobyov,Gov1,Gov2,Gov3}
\begin{equation}
F_P(q^2_1)=-20.69\pm[1.57-2.74][{\rm exp}: 0.48],
\label{eq:FPexp}
\end{equation}
where the first uncertainty bracket includes both the experimental
error as well as all the uncertainties on the input form factors,
the range corresponding again to the range in the $F_A(q^2_1)$ uncertainty,
while the last number in brackets represents the uncertainty following
only from the experimental error in (\ref{eq:exp3He}), namely without any
of the errors on the other inputs. This last number thus indicates the
range of improvement that could be achieved by reducing the uncertainties
on the input form factors $F_V(q^2_1)$, $F_M(q^2_1)$ and $F_A(q^2_1)$,
and especially on the latter one.
Compared to the PCAC expected value for this induced pseudoscalar form factor,
the above result is thus in confirmation of the PCAC prediction with a precision
ranging from 8\% to 13\%. In order to translate this conclusion in terms of
the {\sl nucleon\/} form factor $g_P$, it is necessary to include in the
nuclear model calculation all meson exchange corrections and their associated
uncertainties, thereby leading to a 19\% precise test of PCAC with a value
in agreement with the prediction\cite{Congleton2}. Note also that given
the PCAC relation (\ref{eq:PCAC}), the above value for $F_P(q^2_1)$
inferred from experiment may in turn be used to determine the 
$\pi^\pm-^3$He-$^3$H nuclear coupling constant to much improved
precision\cite{Mukho1}.

A combined fit to two independent observables measured to great
precision, such as $\lambda^{\rm stat}$ and the vector analyzing power
$A_v$, would allow a model independent determination of both $F_A(q^2_1)$
and $F_P(q^2_1)$. However, it may be shown that in order to obtain a result
for $F_A(q^2_1)$ with an uncertainty less than the present value of 0.01,
would require a measurement of $A_v$ to better than 1\%, no small feat indeed!
In particular, when used on its own, 
a 1\% precise measurement of $A_v$, centered onto its theoretical prediction,
would imply the following uncertainty range for $F_P(q^2_1)$ when using 
for the other form factors the values quoted in Sect.\ref{Sect3.1},
\begin{equation}
F_P(q^2_1):\ \ [0.80-0.80]\ [{\rm exp}: 0.55].
\end{equation}

Turning now to the second-class form factors $F_S(q^2_1)$ and $F_T(q^2_1)$,
and using the PCAC prediction (\ref{eq:FPPCAC}) for $F_P(q^2_1)$,
the experimental result (\ref{eq:exp3He}) implies either
\begin{equation}
F_S(q^2_1)=0.026\pm[1.17-2.02]\ [{\rm exp}: 0.38],
\end{equation}
or
\begin{equation}
F_T(q^2_1)=-0.031\pm[1.42-2.45]\ [{\rm exp}: 0.46],
\end{equation}
when either form factor is turned on while the other is still set equal
to zero. Due to the small sensitivity of the statistical capture rate to
these two parameters, these constraints are thus extremely poor, but
nevertheless they improve somewhat the situation existing in terms of the
nucleon form factors $g_S$ and $g_T$\cite{Holstein3} if one is willing
to extrapolate without correction from the hydrogen case.
This conclusion would not be much improved were a 1\% precise measurement
of $A_v$ to become available, since the corresponding uncertainties
ranges are then,
\begin{equation}
F_S(q^2_1):\ \ \pm[0.85-0.89]\ [{\rm exp}: 0.58]\ \ ;\ \ 
F_T(q^2_1):\ \ \pm[0.80-0.83]\ [{\rm exp}: 0.54].
\end{equation}
Similar considerations could of course be developed on the basis of
other observables still, but we shall refrain from doing so since they
are beyond the reach of experiment at present.
  
Finally, let us point out that if $F_P(q^2_1)$ is allowed to vary within
its uncertainty bracket in (\ref{eq:FPexp}), the corresponding values for
$F_S(q^2_1)$ and $F_T(q^2_1)$ then also vary accordingly, but still
within their respective uncertainty brackets. Hence any dependency
of the results for $F_{S,T}$ on the value assumed for $F_P$ is
consistent with their own present uncertainties.

\subsection{Beyond the Standard Model}
\label{Sect3.3}

Let us now turn to the reach for physics beyond the SM offered by the
experimental result (\ref{eq:exp3He}). Defining sensitivities 
$\sigma({\cal O};h_X)$ of observables
to the effective coupling coefficients $h^{S,V,T}_{\pm\pm}$ 
in the same manner as in (\ref{eq:Sen}) with respect to the
vanishing second-class form factors $F_{S,T}$, the corresponding
results are given in Table \ref{Table2}, using as input for nuclear form
factors the values discussed in Sect.\ref{Sect3.1} as well as unit values
for the scalar, pseudoscalar and tensor form factors $G_{S,P,T}$.
In fact, these observables are not sensitive (in linear order) to the
coefficients $h^S_{-\pm}$, $h^T_{-+}$ and $h^V_{+\pm}$ because of the
fact that only a left-handed neutrino couples to muon capture in the
limit of the SM. Note that some of these observables are quite
sensitive to the tensor coupling $h^T_{+-}/2$, for which stringent constraints
will thus be inferred from (\ref{eq:exp3He}).

Indeed, the experimental statistical capture rate (\ref{eq:exp3He}) implies
the following results, turning on only one coupling at a time,
\begin{equation}
\begin{array}{r c l}
h^S_{+-}&=&0.00049\pm[0.0224-0.0385]\ [{\rm exp}: 0.00723],\\
 & & \\
h^S_{++}&=&0.00047\pm[0.0218-0.0374]\ [{\rm exp}: 0.00702],\\
 & & \\
h^S_+\equiv h^S_{+-}+h^S_{++}&=&
0.00048\pm[0.0221-0.0379]\ [{\rm exp}: 0.00712],\\
 & & \\
h^P_+\equiv h^S_{+-}-h^S_{++}&=&-0.0322\pm[1.49-2.55]\ [{\rm exp}: 0.48],\\
 & & \\
\frac{1}{2}h^T_{+-}&=&0.000031\pm[0.00143-0.00245]\ [{\rm exp}: 0.00046],\\
 & & \\
h^V_{--}&=&1.00009\pm[0.00415-0.00712]\ [{\rm exp}: 0.00134],\\
 & & \\
h^V_{-+}&=&-0.000222\pm[0.0102-0.0176]\ [{\rm exp}: 0.0033].
\end{array}
\label{eq:hexp3He}
\end{equation}
These values assume implicitly that the form factors $G_{S,P,T}$ have
been set equal to unity. Note also that the coefficients $h^S_+$ and
$h^P_+$ do define actual scalar and pseudoscalar interactions
at the quark-lepton level, for a neutrino of left-handed chirality.

Again, it may be checked that allowing $F_P$ to vary within its
uncertainty bracket (\ref{eq:FPexp}), each of the above coefficients
then also varies essentially within its own uncertainty bracket.
As could be expected from Table \ref{Table2}, the results for the
scalar couplings $h^S_{+-}$, $h^S_{++}$ and $h^S_+$ 
are already quite stringent,
and in fact improve present limits on such couplings both in the
muonic as well as in electronic sectors. However, the most satisfactory
result is undoubtedly obtained for the tensor coupling
$h^T_{+-}$, which is brought down into the per mille level.
Of course, definite conclusions as to the actual limits for such physics
beyond the SM implied by the experimental result (\ref{eq:exp3He})
would require the evaluation of the form factors $G_S(q^2_1)$, $G_P(q^2_1)$
and $G_T(q^2_1)$. The above limits on $h^{S,V,T}_{\pm\pm}$
are translated in what may be physically more
meaningful terms in Sect.\ref{Sect5}.

Again for comparison, it is interesting to establish what a 1\% precise
measurement of $A_v$, centered onto its theoretical prediction, would
imply for the same set of effective coupling coefficients. Correspondingly,
one finds for the associated uncertainty brackets:
\begin{equation}
\begin{array}{r c l}
h^S_{+-}&:&\pm[0.0164-0.0170]\ [{\rm exp}: 0.0111],\\
 & & \\
h^S_{++}&:&\pm[0.0157-0.0164]\ [{\rm exp}: 0.0107],\\
 & & \\
h^S_+&:&\pm[0.0161-0.0167]\ [{\rm exp}: 0.0109],\\
 & & \\
h^P_+&:&\pm[0.81-0.84]\ [{\rm exp}: 0.55],\\
 & & \\
\frac{1}{2}h^T_{+-}&:&\pm[0.0201-0.0210]\ [{\rm exp}: 0.0136],\\
 & & \\
h^V_{-+}&:&\pm[0.0144-0.0150]\ [{\rm exp}: 0.0098].
\end{array}
\end{equation}
Hence, such results would improve somewhat on the situation  for
some coefficients in (\ref{eq:hexp3He}) following from the experimental 
statistical capture rate (\ref{eq:exp3He}).

\section{The Hydrogen Case}
\label{Sect4}

\subsection{Physical inputs}
\label{Sect4.1}

In the case of muon capture on the proton, the kinematic variables
are such that we have
\begin{equation}
\sqrt{s}=1043.93\ {\rm MeV}\ \ ,\ \ 
\nu=99.146\ {\rm MeV}\ \ ,\ \ 
\omega=944.78\ {\rm MeV}\ \ ,\ \ 
\omega-M_2=5.22\ {\rm MeV}.
\end{equation}
We also use for the CKM $ud$ matrix element the value for $V_{ud}$
quoted in (\ref{eq:Vud}), including its uncertainty. Indeed, the latter
contributes 0.16\% to the final uncertainty on the theoretical prediction,
and should thus be included given the aim of a 0.5\% precise measurement
of the singlet capture rate\cite{muH,Vorobyov}.
As to the overlap reduction factor $C$, the value used is that
detailed in Appendix 1 in the case of the proton, namely $C=0.9956$.

Let us turn to the issue of the nucleon electroweak form factors,
to be evaluated at the momentum transfer $q^2_0=-0.877 m^2_\mu$,
beginning with the vector ones, $g_V(q^2_0)$ and $g_M(q^2_0)$. Since
through CVC, the values for these nuclear form factors are related
to the electromagnetic ones for the proton and the neutron
whose charge radii are very well known\cite{Mergell}, it becomes possible
to infer very precise values for $g_V(q^2_0)$ and $g_M(q^2_0)$. From
the values $r^v_1=0.765(1\pm0.01)$ fm and $r^v_2=0.893(1\pm 0.01)$ fm
(see Ref.\cite{Mergell} for the meaning of these parameters), as well as
the proton and neutron anomalous magnetic moments and the fact that
$g_V(q^2=0)=1$, the following values apply
\begin{equation}
g_V(q^2_0)=0.9755\pm 0.0005\ \ ,\ \ 
g_M(q^2_0)=3.5821\pm 0.0025.
\end{equation}

The value for the axial form factor $g_A(q^2_0)$ is inferred from
the neutron decay rate, which implies\cite{PDG},
\begin{equation}
g_A(q^2=0)=1.2670\pm 0.0035,
\end{equation}
as well as the mean square axial charge 
radius\cite{Bernard2,Bernard1,Bernard3,Bernard4}
\begin{equation}
r_A=0.65\pm 0.03\ {\rm fm}.
\end{equation}
Consequently, one finds
\begin{equation}
g_A(q^2_0)=1.245\pm 0.004.
\end{equation}
Note that since the capture rates are rather sensitive to the axial form
factor (see below), if the value for $g_A(q^2=0)$ were to change
again, the values for capture rates would have to be adapted
appropriately.

The value for the induced pseudoscalar
form factor $g_P(q^2_0)$ is given by the prediction based on the chiral
symmetries of QCD\cite{Bernard1,Fearing,Bernard3,Bernard4},
\begin{equation}
g^\chi_P(q^2_0)=\frac{2m_\mu f_\pi g_{\pi NN}}{m^2_\pi-q^2_0}-
\frac{1}{3}g_A(0)m_\mu M r^2_A,
\end{equation}
where $f_\pi=92.5\pm 0.2$ MeV is the usual pion decay constant
and $g_{\pi NN}$ the pion-nucleon coupling constant. The latter
quantity has been the issue of much debate, but using the recent
precise estimate of Ref.\cite{Loiseau},
\begin{equation}
g_{\pi NN}=13.37\pm 0.09,
\end{equation}
one finds
\begin{equation}
g^\chi_P(q^2_0)=8.475\pm 0.076,
\label{eq:gPchiral}
\end{equation}
namely a chiral symmetry prediction with a precision better than 1\%.

The second-class form factors $g_S$ and $g_T$ are taken to
be identically vanishing, an appro\-xi\-ma\-tion discussed in Sect.\ref{Sect2.2}
which is justified in the limit of exact isospin symmetry. 
However, the effects of isospin breaking are estimated to be 
small\cite{Donoghue,Holstein2,Dominguez,Shiomi}, on the order of
0.02 for $|g_S/g_V|$ and $|g_T/g_A|$, and given the sensitivity of the
capture rates to these two form factors (see below), 
their contribution may safely be neglected.

Finally, there is the issue of the nucleon scalar, pseudoscalar and tensor
form factors $G_S(q^2_0)$, $G_P(q^2_0)$ and $G_T(q^2_0)$, whose values
are unknown at present. Any estimate would of course be welcome,
but the expected result should be of the order of unity, within may be a
factor of at most ten, as mentioned previously.

\subsection{Within the Standard Model}
\label{Sect4.2}

Given the different inputs discussed in Sect.\ref{Sect4.1}, it is 
straightforward to consider predictions for the observables in muon
capture on the proton within the SM. Setting all effective coupling
coefficients $h^{S,V,T}_{\pm\pm}$ to zero except for $h^V_{--}=1$, one
then finds,
\begin{equation}
\begin{array}{r c l}
\lambda^{\rm S}&=&688.4\pm 3.8\ {\rm s}^{-1},\\
\lambda^{\rm T}&=&12.01\pm 0.12\ {\rm s}^{-1},\\
\lambda^{\rm stat}&=&181.11\pm 0.98\ {\rm s}^{-1},\\
A_\Delta&=&-0.9337\pm 0.0007,\\
A_v&=&0.00371\pm 0.00030,\\
A_t&=&-0.06260\pm 0.00052.
\end{array}
\end{equation}
Note the small values for the latter two observables, rendering their
experimental determination essentially impossible, given the small
polarizations available for the muonic hydrogen atom.

As in the case of muon capture on $^3$He, it is interesting to consider
the sensitivity of each of these observables to the different input
nucleon form factors, in particular $g_P(q^2_0)$. These sensitivities
are defined as in (\ref{eq:Sen}), and their values are presented
in Table \ref{Table3}. Note that with respect to the measured
capture rates, namely the statistical one for $^3$He and the singlet one
for the proton, their sensitivity to the induced pseudoscalar form factor
is essentially comparable, as is also the case for the other nucleon
form factors.

In fact, assuming the goal reached\cite{muH,Vorobyov} of measuring the
singlet rate on the proton to a precision of 0.5\% with a central value
equal to the above theoretical prediction precise to 0.55\%, one may
infer the following value for the induced pseudoscalar form factor,
\begin{equation}
g_P(q^2_0)=g^\chi_P(q^2_0)\pm 0.327\ [{\rm exp}: 0.230],
\end{equation} where the second indicated uncertainty follows only
from the experimental precision of 0.5\%, while the first also includes
all the errors on the theoretical inputs for the other form factors
and the CKM matrix element $V_{ud}$. In other words, a 0.5\% precise
measurement of the singlet capture rate implies a 3.9\% precise
determination of $g_P$ and for the test of the chiral symmetry prediction 
of that value. In the same way as in Ref.\cite{Mukho1}, the same result may
also be used to infer a value for the pion-nucleon coupling constant
$g_{\pi NN}$, assuming that $g_P(q^2_0)$ takes its expected value
(\ref{eq:gPchiral}),
\begin{equation}
g_{\pi NN}=13.37\pm 0.49,
\end{equation}
hence a 3.7\% precise determination of that quantity.

Pursuing along the same lines, assuming the value for $g_P(q^2_0)$ set
by the chiral symmetry prediction, and a measurement of the singlet rate
precise to 0.5\% in the manner described above, each of the second-class
form factors may also be constrained to the following precision,
\begin{equation}
g_S:\ \ \pm 0.314\ [{\rm exp}: 0.216]\ \ ,\ \ 
g_T:\ \ \pm 0.307\ [{\rm exp}: 0.210], 
\end{equation}
thus providing much of an improvement on the present 
situation\cite{Holstein3}, but still an order of magnitude away from
the expected range of values.

\subsection{Beyond the Standard Model}
\label{Sect4.3}

The sensitivities of the considered observables to the effective coupling
coefficients $h^{S,V,T}_{\pm\pm}$ are given in Table \ref{Table4},
which displays some interesting differences with the $^3$He case,
but again not between the singlet capture rate on the proton and the
statistical rate for $^3$He.

In order to assess the potential reach offered by the singlet rate for
physics beyond the SM, let us again assume a 0.5\% precise measurement
of that observable centered onto its theoretical prediction. One then finds
the following uncertainties,
for each of the relevant effective coupling coefficients 
$h^{S,V,T}_{\pm\pm}$, turning them on each one after the other,
\begin{equation}
\begin{array}{r c l}
h^S_{+-}&:&\ \ \pm 0.0168\ [{\rm exp}: 0.0115],\\
 & & \\
h^S_{++}&:&\ \ \pm 0.0187\ [{\rm exp}: 0.0128],\\
 & & \\
h^S_+=h^S_{+-}+h^S_{++}&:&\ \ \pm 0.0177\ [{\rm exp}: 0.0121],\\
 & & \\
h^P_+=h^S_{+-}-h^S_{++}&:&\ \ \pm 0.336\ [{\rm exp}: 0.230],\\
 & & \\
\frac{1}{2}h^T_{+-}&:&\ \ \pm 0.00140\ [{\rm exp}: 0.00096],\\
 & & \\
h^V_{--}&:&\ \ \pm 0.00364\ [{\rm exp}: 0.00250],\\
 & & \\
h^V_{-+}&:&\ \ \pm 0.0113\ [{\rm exp}: 0.0065],
\end{array}
\end{equation}
in which it is implicitly assumed that the form factors
$G_{S,P,T}$ have all three been set to a unit value.
Consequently, given the results in (\ref{eq:hexp3He}),
the physics reach offered by the singlet muon capture rate
on the proton measured to 0.5\% precision improves by factors of at
least two to three the results achieved
already from the available statistical capture rate on $^3$He.

\section{Specific Models Beyond the Standard Model}
\label{Sect5}

The effective interaction (\ref{eq:effec}) in terms of the coupling
coefficients $h^{S,V,T}_{\pm\pm}$ provides a model independent parametrization
for any low-energy contributions stemming from whatever new physics there
is lurking behind the confines of the Standard Model. However, taken
as such, the constraints (\ref{eq:hexp3He}) obtained from the experimental
statistical capture on $^3$He probably do not mean much in terms of
the energy scales or coupling strengths implied by such constraints.
To develop an idea for the latter kind of data however, it becomes
necessary to consider specific models for physics beyond the SM. Three
general such classes will briefly be considered here, with the notations
defined in Appendix 3. Other possibilities come to mind, such as
for example models with large extra dimensions which may be amusing to 
investigate in the same vain. Also, when the singlet capture rate on the
proton will have been measured, similar considerations may be developed
as well, improving to some extent on the present results.

\subsection{Left-Right symmetric models}

Turning first to left-right symmetric models (LRSM), only the limits
on the couplings $h^V_{--}$ and $h^V_{-+}$ need to be considered, since
only vector and axial interactions are implied in such models (up to small
scalar Higgs exchange interactions which may be ignored, see Appendix 3). 
Given the value from (\ref{eq:hexp3He}),
\begin{equation}
h^V_{--}=1.00009\pm 0.00712,
\end{equation}
different considerations may be developed. First, one may view this result
as a constraint on the universality of the electroweak 
interactions\cite{Mukho},
whose most stringent limit in the electron-muon flavour sector follows
from $\pi^\pm$ decays\cite{Herczeg,PDG} at the 0.4\% level. The above limit
implies a universality constraint at the 1.4\% level only (in terms of
$(h^V_{--})^2$).

The above result for $h^V_{--}$ may also be viewed as a constraint
on the unitarity properties of the CKM quark flavour mixing matrix,
in terms of the ratio $V_{ud}/V^{\rm unitary}_{ud}$. Indeed,
in muon capture one is dealing entirely with the muon flavour sector
only, while $V_{ud}$ usually involves different $\beta$-decay processes,
hence the electron flavour sector. Therefore, another way to read the
constraint on $h^V_{--}$ is to say that the experimental statistical
capture rate on $^3$He confirms the unitary-constrained value for
$V_{ud}$, to a level better than what is achieved in terms of the
$0^+$-$0^+$ superallowed $\beta$-decays where conclusions are
somewhat dependent on the nuclear models used to evaluate the 
radiative corrections\cite{PDG}.

Finally, within LRSM, the above result for $h^V_{--}$ does not imply
any limit on the mass $M^W_2$ of the extra charged gauge boson, since
an expression similar to that given in Appendix 3 for the coefficient
$h^V_{--}$ also applies to the similar coefficient $f^V_{--}$ relevant
to $\beta$-decay, so that when reexpressing all couplings in terms of
physical quantities (such as the muon decay rate, and so on), such factors
essentially cancel. As a matter of fact, the sole contribution which survives
this comparison of muon decay, $\beta$-decay and muon capture amplitudes,
is that in LRSM the CKM leptonic flavour mixing matrices may be different
for the electron and muon flavours. Thus in fact, the above limit
on $h^V_{--}$ translates into the following constraint on leptonic
flavour mixing in these two sectors, when the mixing angle $\zeta$ of the
charged gauge bosons is ignored, $\zeta=0$\cite{Gov1},
\begin{equation}
(v_\mu-v_e)v_ur^4\delta^2=-0.00018\pm 0.01424,
\end{equation}
with
\begin{equation}
v_\mu=\frac{\sum_i'|U^R_{\mu i}|^2}{\sum_i'|U^L_{\mu i}|^2}\ \ ,\ \ 
v_e=\frac{\sum_i'|U^R_{e i}|^2}{\sum_i'|U^L_{e i}|^2}\ \ ,\ \ 
v_u=\frac{|V^R_{ud}|^2}{|V^L_{ud}|^2},
\end{equation}
in which the summation over the index $i$ stands for all neutrino mass
eigenstates whose production is kinematically allowed in muon capture
on the one hand, and in $\beta$-decay on the other (in which case,
their mass is taken to be negligeable as well).
The other parameters are defined in Appendix 3.

Similarly, the result on $h^V_{-+}$ in (\ref{eq:hexp3He}) may be
translated in terms of parameters of LRSM as (see Appendix 3),
\begin{equation}
rt\,{\rm Re}\left(e^{i\omega} v_{ud}\right)=0.000222\pm 0.0176.
\end{equation}
In the case of manifest LRSM with $r=1$, $v_{ud}=1$ and $\omega=0$,
this result provides a limit on the mixing angle $\zeta$
which does not improve such limits stemming already from $\beta$-decay
processes\cite{PDG}.

In fact, again in the limit of a vanishing mixing angle $\zeta$, the
vector analyzing power $A_v$ reduces to
\begin{equation}
{A^{\rm LRSM}_v}_{|_{\zeta=0}}
=\frac{1-r^4\delta^2 v_\mu v_u}{1+r^4\delta^2 v_\mu v_u}\,
A^{\rm SM}_v,
\end{equation}
with $A^{\rm SM}_v$ its SM value. However, even a measurement to
1\% of $A_v$ would not imply a lower bound on $M^W_2$ better than
$260$ GeV (95\% C.L.) in the manifest LRSM.

\subsection{Contact interactions}

Let us now consider the possibility of contact interactions (see Appendix 3).
Given the pa\-ra\-me\-tri\-za\-tion of such interactions, it is a simple matter
to translate the limits (\ref{eq:hexp3He}) in terms of the associated
compositeness scales, with the following lower bounds (in an obvious notation)
all given at the 95\% C.L.,
\begin{equation}
\begin{array}{r c l}
\Lambda^S_{+-}&>&1.60\ {\rm TeV},\\
& & \\
\Lambda^S_{++}&>&1.62\ {\rm TeV},\\
& & \\
\Lambda^S_{+}&>&1.61\ {\rm TeV},\\
& & \\
\Lambda^P_{+}&>&196\ {\rm GeV},\\
& & \\
\Lambda^T_{+-}&>&6.34\ {\rm TeV},\\
& & \\
\Lambda^V_{+-}&>&2.36\ {\rm TeV}.
\end{array}
\end{equation}
Clearly, the limits on $\Lambda^V_{+-}$ and $\Lambda^T_{+-}$
are quite competitive with recent
collider results in some of these channels\cite{H1}. One also has to
keep in mind that the latter results apply to the electron sector,
while those established here on basis of the experimental statistical
capture rate on $^3$He apply to the muon sector for couplings between
the first quark generation and the second lepton one. On the other hand,
the above lower bounds are not as stringent as those which follow from
atomic parity violation\cite{Barger}, but again the latter limits apply to the
electron sector and for neutral current interactions.
It is thus fair to say that precision studies in nuclear muon capture
on simple nuclei have at present the potential to test the SM in
sectors and in ways complementary to those accessible through collider
experiments, hence the physics interest of the limits established above.

\subsection{Leptoquarks}

Finally, let us consider the possibility of 
leptoquarks (LQ)\cite{Buchmuller,Davinson}, in the
notations introduced in Appendix 3. Among all the possible limits
which may be obtained from (\ref{eq:hexp3He}), only the most
stringent ones are presented here.

In the case of scalar LQ, the limit follows from the value for
the tensor coupling coefficient $h^T_{+-}$, leading to
\begin{equation}
\frac{M}{\lambda}>894\ {\rm GeV}\ \ (95\%\ {\rm C.L.}).
\end{equation}
The notation used here is symbolic. Indeed, referring back to
the expressions for the $h^{S,V,T}_{\pm\pm}$ coefficients in terms of
the LQ couplings and parameters, one sees that the $h^T_{+-}$ coupling
involves either the $S_0(1/3)$ LQ with the couplings 
$\lambda^{L*}_{S_0}\lambda^R_{S_0}$, or the $S_{1/2}(-2/3)$ LQ with the
couplings $\lambda^{L*}_{S_{1/2}}\lambda^R_{S_{1/2}}$. Thus in the above
lower bound, $M$ stands for the mass of one or the other of these scalar LQ,
while $\lambda$ stands for the square root of the product of the two
associated couplings constants.

Similarly considering the limit in (\ref{eq:hexp3He}) on the coefficient
$h^S_{++}$, one finds the lower bound,
\begin{equation}
\frac{M}{\lambda}>915\ {\rm GeV}\ \ (95\%\ {\rm C.L.}),
\end{equation} 
in a similar notation referring now to either vector LQ
$V_0(-2/3)$ or $\tilde{V}_{1/2}(-1/3)$ (for the associated combination
of LQ couplings corresponding to the factor $\lambda$, see Appendix 3).
Again, these limits are certainly as stringent as those recently presented
in Ref.\cite{H1,D0}. This is particularly relevant when one recalls again that
the constraints from Ref.\cite{H1} refer to the electron sector, while
the limits established here apply to LQ coupling the first quark generation
to the second lepton generation.

\section{Conclusions}
\label{Sect6}

This paper provides for the first time explicit and complete 
analytic expressions for all
obser\-va\-bles relevant to nuclear muon capture on a spin 1/2 isospin doublet,
thus of direct use to the cases of the proton and $^3$He. The results include
all possible contributions for the nuclear matrix elements, as well as
all possible effective interactions beyond the usual electroweak charged
interaction. The analysis was developed with great care, keeping 
approximations to the strictest minimum and in ways not affecting the
final numerical results, including the
calculation of nuclear finite size corrections to the muon overlap
correction factor, and considering the limit of a Dirac neutrino of zero mass. 
Such expressions are ideally suited for tests
of the Standard Model from precision muon capture experiments,
both in addressing still open questions related to the strong nonperturbative
sector of the quark interactions probed through the electroweak sector,
as well as in probing for any new physics which is lying in store just 
waiting to be discovered experimentally.

These results also enable specific predictions for observables with 
a precision on a par with the challenge on the experimental side as well.
Specific new results for muon capture on the proton have been given,
and some others in the case of $^3$He. This situation is particularly
relevant given the recent precision measurement of the statistical
capture rate on $^3$He\cite{Ackerbauer}, and the projected 0.5\% precise
measurement of the singlet rate on the proton\cite{muH,Vorobyov}.

The discussion showed how stringent limits can be inferred from the
$^3$He result already, both within the SM and beyond it, sometimes
in real competition with results from collider experiments, and often
complementary to these. The physics reach of the foreseen experiment
on the proton has also been assessed. In due time, when that experiment
will have been completed to the desired level of precision,
similar but improved limits and tests on the SM will be inferred, and 
in particular the issue of the induced pseudoscalar nucleon form factor 
settled once and for all, presumably in perfect agreement with the 
theoretical expectation. Also, the potential physics reach of other 
observables may now be assessed completely on the basis of the expressions 
of this paper, even though the actual measurement of any of these observables, 
which all involve either initial polarization muonic atom states or final 
polarization measurements, or both, poses an almost impossible experimental 
challenge. But such a situation has never deterred any experimentalist at 
heart, quite to the contrary! We hope that this paper will also be of use in
the experimental pursuit of the impossible polarization observables.

\section{Acknowledgments}

One of us (JG) wishes to thank Tom Case, Jules Deutsch, Peter Kammel, Claude
Petitjean and Alexei Vorobyov for the many conversations we have had on
the topics of this work over the last years. This work has been completed while
each of the authors is enjoying a sabbatical leave from his home
Institution. We wish to thank for their warm hospitality and support
our present host Institutions, Peter van Nieuwenhuizen and
the C.N. Yang Institute for Theoretical Physics of the State University of 
New York at Stony Brook (USA), and the INFN National Laboratory 
at Frascati (Italy).  

This work has been partially supported under the terms of an agreement
between the CONACyT (Mexico) and the FNRS (Belgium).

\clearpage

\section*{Appendix 1: The Overlap Reduction Factor}
\label{A1}

In the calculation as outlined in Section 2, it is implicitly
assumed that one is dealing with point-like particles whose quantum
states are described by plane waves. Since this is clearly not the case,
further corrections have to be introduced in order to account for
the finite spatial extent of the nuclear charge distributions,
for the bound state character of the initial muonic atom, and
for possible relativistic corrections since it is the non relativistic
Coulomb wave function $\psi_c(r)$ which is initially considered to
represent the muon probability amplitude rather than the full-fledged
solution to the relevant Dirac equation. Indeed, all these effects
ought to be carefully assessed in order to obtain a trustworthy
evaluation of the reduction factor $C$ introduced in (\ref{eq:kinerate}),
with a precision at the required level with respect to experimental aims.

This Appendix presents such an evaluation, following closely the discussion
developed in Ref.\cite{JimPhD} in the case of $^3$He. Since such an
analysis, which is important for our purposes, is not available in the
literature, we felt it useful to include it here, the more so since it
seems to have been the intention of the author of Ref.\cite{JimPhD} to
make it available.

Nuclear charge distributions, whether of electric or electroweak matter,
need to be modeled in such an analysis. This we shall do in terms of
a spherically symmetric density $\rho_m(r)$, where the index distinguishes
the different types of matter encountered in the problem, normalized
to unity over the volume of space,
\begin{equation}
\int_{(\infty)}d^3\vec{r}\,\rho_m(r)=1,
\end{equation}
and dependent on a single length scale parameter $a_m$ directly related to the
mean square radius of the distribution. The simplest such model which we
shall use is of the form
\begin{equation}
\rho_m(r)=\frac{1}{8\pi a^3_m}e^{-r/a_m},
\label{eq:chargedistri}
\end{equation}
such that
\begin{equation}
r^2_m\equiv <r^2>_m=\int_{(\infty)}d^3\vec{r}\,r^2\rho_m(r)=12a^2_m.
\end{equation}
Other charge distribution models may be considered of course\cite{JimPhD},
leading to no significant difference in the evaluation of the reduction
factor $C$. Note that the form factor in momentum space associated to the
model (\ref{eq:chargedistri}) is of the usual dipolar form,
\begin{equation}
F_m(q^2)=\frac{1}{(1-a^2_mq^2)^2}.
\end{equation}

Given such models, the reduction factor $C_m$ associated to each of 
these nuclear matter distributions is thus given by\cite{JimPhD}
\begin{equation}
\sqrt{C_m}=\int_{(\infty)}d^3\vec{r}\,(1+\nu^2a^2_m)^2\rho_m(r)\,
j_0(\nu r)\,\frac{\psi_1(r)}{\psi^{(0)}_1(0)},
\label{eq:reduction}
\end{equation}
where $\nu$ stands for the neutrino energy. In this expression,
the factor $(1+\nu^2a^2_m)^2$ stems from the normalization of the
form factor $F_m(q^2)$ or the normalization condition for the
distribution $\rho_m(r)$, $j_0(\nu r)=\sin(\nu r)/(\nu r)$ 
is the spherical Bessel function associated to the angle-integrated
neutrino plane wave function $e^{-i\vec{p}\cdot\vec{r}}$,
$\psi_1(r)$ is the ground state wave function of the muonic atom,
and finally $\psi^{(0)}_1(r)=\psi_c(r)$ is the 1S ground state Coulomb
wave function of the muonic atom since this specific choice was made
in the normalization of the capture distribution (\ref{eq:kinerate}).
Hence, only the function $\psi_1(r)$ is left to be computed in order
to evaluate the above overlap integral giving the reduction factor
$C_m$.

The wave function $\psi_1(r)$ is that of the muon (``carrying" the reduced
mass $\mu$ of the muonic atom) in the electrostatic field of the nucleus
of finite size. In principle, this function is to be obtained by solving
the associated Dirac equation in the given electrostatic potential,
but one would expect that the ensuing relativistic corrections should be
sufficiently small to be ignored since of order $(\alpha Z)^2$, which is 
the squared velocity of the bound muon. This expectation is indeed borne
out by the detailed numerical resolution of the associated Dirac
equation in the case of $^3$He\cite{JimPhD} and of the proton, 
and this relativistic correction shall thus not 
be included here in the evaluation of $C_m$.

Hence, this leaves only to solve the Schr\"odinger equation in the
same electrostatic potential of the electric charge distribution of the
initial nucleus. For the latter distribution, we shall again use a model
of the form (\ref{eq:chargedistri}), with a parameter $a_c$ related to
the electric charge mean square radius of that nucleus. Correspondingly,
in addition to the usual Coulomb potential for a point charge of value
$(+Ze)$, one needs to include in the Schr\"odinger equation the following
perturbation in the potential\cite{JimPhD},
\begin{equation}
\Delta V(r)=\alpha Z\hbar c\left(\frac{1}{r}+\frac{1}{2a_c}\right)
e^{-r/a_c}.
\label{eq:potential}
\end{equation}
The contributions of this term to the 1S ground state wave function 
$\psi_1(r)$ are then computable through perturbation theory. It
appears that the first order correction is already sufficient for
our purposes since it is proportional to the factor 
$(a_c/a_0)^2\sim 10^{-5}$, where $a_0$ is the usual atomic Bohr radius
{\sl for the muonic atom\/}, thus given by
\begin{equation}
a_0=\frac{1}{\alpha Z}\frac{\hbar c}{\mu c^2}.
\end{equation}

The spherically symmetric solutions to the Coulomb problem are well
known. For the $nS$ state, $n=1,2,\cdots$, one has
\begin{equation}
\psi^{(0)}_n(r)=\frac{1}{\sqrt{\pi a^3_0n^5}}\,e^{-r/(na_0)}
L^1_{n-1}\left(\frac{2r}{na_0}\right)\ \ ,\ \ 
\psi_c(0)\equiv\psi^{(0)}_1(0)=\frac{1}{\sqrt{\pi a^3_0}},
\end{equation}
where $L^1_{n-1}(x)$ are the usual Laguerre polynomials. The
associated (binding) energy eigenvalues are
\begin{equation}
E^{(0)}_n=-\frac{1}{2}(\alpha Z)^2\mu c^2\,\frac{1}{n^2}.
\end{equation}
Adding the correction (\ref{eq:potential}), first-order perturbation
theory then leads to the following $1S$ ground state normalized
wave function,
\begin{equation}
\psi_1(r)=\psi^{(0)}_1(r)-16\left(\frac{a_c}{a_0}\right)^2
\sum_{n=2}^\infty\frac{1}{\sqrt{n^3}}\frac{1}{1-\frac{1}{n^2}}
\,\psi^{(0)}_n(r),
\label{eq:correctwave}
\end{equation}
while the associated energy is
\begin{equation}
E_1=-\frac{1}{2}(\alpha Z)^2\mu c^2
\left(1-16\left(\frac{a_c}{a_0}\right)^2\right).
\end{equation}

The evaluation of the overlap reduction factor $C_m$ in (\ref{eq:reduction})
is now straightforward using the integrals
\begin{equation}
\frac{1}{\psi^{(0)}_1(0)}\int_{(\infty)}d^3\vec{r}\,
\rho_m(r)\,\psi^{(0)}_n(r)=\frac{1}{\sqrt{n^3}}\,
\frac{\left(1-\frac{1}{n}\frac{a_m}{a_0}\right)^{n-2}}
{\left(1+\frac{1}{n}\frac{a_m}{a_0}\right)^{n+2}}
\left(1-\frac{a_m}{a_0}\right).
\end{equation} 
One then finds
\begin{equation}
\sqrt{C_m}=\frac{1}{\left(1+\frac{a_m}{a_0}\right)^3}
\left(\frac{1+\nu^2a^2_m}{1+\frac{\nu^2a^2_m}
{\left(1+\frac{a_m}{a_0}\right)^2}}\right)^2-
4\left(\frac{a_c}{a_0}\right)^2,
\end{equation}
where in the second term which follows from the first-order
correction in (\ref{eq:correctwave}) the limit $a_m\rightarrow 0$
has been taken. Numerically, this result does not differ from the
same expression expanded to second order\footnote{Both these ratios
being on the order of $10^{-3}$, cubic corrections are indeed totally
negligeable for our purposes.} in $a_c/a_0$ and $a_m/a_0$\cite{JimPhD},
\begin{equation}
\sqrt{C_m}=1-3\frac{1-\frac{1}{3}\nu^2a^2_m}{1+\nu^2a^2_m}
\left(\frac{a_m}{a_0}\right)
+6\frac{1-\nu^2a^2_m}{(1+\nu^2a^2_m)^2}\left(\frac{a_m}{a_0}\right)^2
-4\left(\frac{a_c}{a_0}\right)^2.
\end{equation}

Let us now turn to numerical evaluations, first in the case of $^3$He.
The values for the relevant parameters $a_m$ may only be inferred
from a nuclear model calculation of the corresponding mean square 
charge radii. This is done in Ref.\cite{JimPhD} with the following results,
\begin{equation}
a_{(1)}=0.554\ {\rm fm}\ \ ,\ \ 
a_{(\sigma)}=0.512\ {\rm fm},
\end{equation}
where $a_{(1)}$ (resp. $a_{(\sigma)}$) stands for the parameter $a_m$
associated to form factors for the vector (resp. axial) current
(thus including the electromagnetic current in the vector case).
Using then the mass values and neutrino energy given in Section \ref{Sect3.1},
one finds
\begin{equation}
\frac{a_c}{a_0}=4.17\cdot 10^{-3}\ \ ,\ \ 
C_{(1)}=0.9777\ \ ,\ \ 
C_{(\sigma)}=0.9790.
\end{equation}
Weighing these two reduction factors with the relative vector and
axial current contributions to the capture rates, an effective value
of $C_{\rm eff}=0.9788$ is obtained, thus finally leading to the
overlap reduction factor for $^3$He as determined in Ref.\cite{JimPhD},
\begin{equation}
C(^3{\rm He})=0.979.
\end{equation}

In the case of the proton, things may be done with great precision
since a great deal is known about the nucleon electromagnetic form
factors\cite{Mergell,Bernard2,Bernard1}. Given the value 
$r^p_E=0.847$~fm\cite{Mergell} (known to 1\% precision)
for the proton electric charge distribution, one has
\begin{equation}
\frac{a_c}{a_0}=8.59\cdot 10^{-4}.
\end{equation}
On the other hand, the values for $C_{V,M,A}$ may be determined using
the associated charge radii $r^1_v$, $r^2_v$ and $r_A$ discussed in
Section \ref{Sect4.1}, including their uncertainties. One then finds,
\begin{equation}
C_V=0.99534\pm 0.00004\ \ ,\ \ 
C_M=0.99460\pm 0.00005\ \ ,\ \ 
C_A=0.9960\pm 0.0002.
\end{equation}
Assuming then that the correction factor associated to the
nuclear matter distribution charac\-te\-ri\-zed by the induced pseudoscalar
form factor $g_P$ is also given by $C_A$, and weighing each of these
correction factors with the respective relative contributions 
of the associated form factors to the capture rates,
the following effective value for the overlap reduction factor is
finally obtained,
\begin{equation}
C(p)=0.9956.
\end{equation}
Note that this value agrees essentially with the 0.4\% correction
quoted in Ref.\cite{Santisteban} in the case of hydrogen.
Here however, a careful assessment of all possible effects has led
to this final value, with increased precision.

\section*{Appendix 2: Polarization and Hyperfine States}
\label{A2}

In the calculation outlined in Sect.\ref{Sect2}, the natural
representation of polarization states is through the normalized spin vectors
$\hat{s}_{\mu,1,2}$ rather than the hyperfine states $(S=1,m=0,\pm 1)$
and $(S=0,m=0)$ of the muonic atom. However, there exists
a correspondence between these two bases for the initial spin degrees
of freedom, and the purpose of this Appendix is to indicate how
results pertaining to these hyperfine states may be extracted from the
results derived in Sect.\ref{Sect2}.

First, let us note that for a spin 1/2 system, any of its normalized
quantum states may be parametrized as
\begin{equation}
|\hat{s}>=e^{-i\varphi/2}\cos\frac{\theta}{2}\,|+>\,+\,
e^{i\varphi/2}\sin\frac{\theta}{2}\,|->,
\end{equation}
where $|\pm>$ represent the basis of states with spin eigenvalues
$m=\pm 1/2$ with respect to an arbitrary quantization axis.
The angular variables $(\theta,\varphi)$ may be interpreted as being 
the spherical coordinates for a unit vector $\hat{s}$ whose components are 
defined as in (\ref{eq:spherical}), since one finds for the associated
spin component operators given in terms of the usual Pauli matrices
$\vec{\sigma}$,
\begin{equation}
<\vec{\sigma}>\equiv<\hat{s}|\vec{\sigma}|\hat{s}>=\hat{s}
\end{equation}
(note also that the state $|\hat{s}>$ corresponds to the bi-spinor
$\chi_+(\hat{s})$ in (\ref{eq:chifunctions})).

Let us now consider the quantum spin states of a bound system of
two spin 1/2 particles, such as the muonic atom, obtained through
the tensor product $|\hat{s}_\mu>_\mu|\hat{s}_1>_1$
of the associated spin states,
\begin{equation}
|\hat{s}_\mu>_\mu=e^{-i\varphi_\mu/2}\cos\frac{\theta_\mu}{2}\,|+>_\mu\,+\,
e^{i\varphi_\mu/2}\sin\frac{\theta_\mu}{2}\,|->_\mu,
\end{equation}
\begin{equation}
|\hat{s}_1>_1=e^{-i\varphi_1/2}\cos\frac{\theta_1}{2}\,|+>_1\,+\,
e^{i\varphi_1/2}\sin\frac{\theta_1}{2}\,|->_1.
\end{equation}
In terms of the usual $|S,m>$ hyperfine state
\begin{equation}
|1,1>=|+>_\mu|+>_1\ \ ,\ \ 
|1,-1>=|->_\mu|->_1, 
\end{equation}
\begin{equation}
|1,0>=\frac{1}{\sqrt{2}}\left[|+>_\mu|->_1+|->_\mu|+>_1\right]\ \ ,\ \ 
|0,0>=\frac{1}{\sqrt{2}}\left[|+>_\mu|->_1-|->_\mu|+>_1\right],
\end{equation}
one then finds
\begin{equation}
|\hat{s}_\mu>_\mu|\hat{s}_1>_1=
A_{1,1}|1,1>+A_{1,-1}|1,-1>+A_{1,0}|1,0>+A_{0,0}|0,0>,
\end{equation}
where
\begin{equation}
A_{1,1}=e^{-i(\varphi_\mu+\varphi_1)/2}\,
c_\mu c_1\ \ ,\ \ 
A_{1,-1}=e^{i(\varphi_\mu+\varphi_1)/2}\,
s_\mu s_1,
\end{equation}
\begin{equation}
A_{1,0}=\frac{1}{\sqrt{2}}\left[e^{-i(\varphi_\mu-\varphi_1)/2}
c_\mu s_1\,+\,
e^{i(\varphi_\mu-\varphi_1)/2}
s_\mu c_1\right], 
\end{equation}
\begin{equation}
A_{0,0}=\frac{1}{\sqrt{2}}\left[e^{-i(\varphi_\mu-\varphi_1)/2}
c_\mu s_1\,-\,
e^{i(\varphi_\mu-\varphi_1)/2}
s_\mu c_1\right],
\end{equation}
with $c_\mu=\cos\theta_\mu/2$, $s_\mu=\sin\theta_\mu/2$, $c_1=\cos\theta_1/2$
and $s_1=\sin\theta_1/2$.
In particular, the hyperfine populations are then given as
\begin{equation}
N_{1,1}=\frac{1}{4}\left[1+\cos\theta_\mu\right]\left[1+\cos\theta_1\right]
\ \ ,\ \ 
N_{1,-1}=\frac{1}{4}\left[1-\cos\theta_\mu\right]\left[1-\cos\theta_1\right],
\end{equation}
\begin{equation}
N_{1,0}=\frac{1}{4}\left[1-\cos\theta_\mu\cos\theta_1+
\cos(\varphi_\mu-\varphi_1)\sin\theta_\mu\sin\theta_1\right], 
\end{equation}
\begin{equation}
N_{0,0}=\frac{1}{4}\left[1-\cos\theta_\mu\cos\theta_1-
\cos(\varphi_\mu-\varphi_1)\sin\theta_\mu\sin\theta_1\right], 
\end{equation}
with $N_{1,1}+N_{1,-1}+N_{1,0}+N_{0,0}=1$, as it should.

Let us now consider the matrix element 
${\cal M}_\lambda(\hat{s}_\mu,\hat{s}_1)=
<\hat{s}_2,\lambda|\hat{H}_{\rm eff}|\hat{s}_\mu,\hat{s}_1>$
which was expressed in Sect.\ref{Sect2.2} (in a notation which should be
self-explanatory). In terms of the above change of basis in spin space,
one then finds the following decomposition into hyperfine capture
amplitudes ${\cal M}^{S,m}_\lambda$,
\begin{equation}
{\cal M}_\lambda(\hat{s}_\mu,\hat{s}_1)=A_{1,1}{\cal M}^{1,1}_\lambda+
A_{1,-1}{\cal M}^{1,-1}_\lambda+
A_{1,0}{\cal M}^{1,0}_\lambda+
A_{0,0}{\cal M}^{0,0}_\lambda.
\end{equation}
The hyperfine capture distributions are then given as in (\ref{eq:kinerate})
with the quantity $|{\cal M}_\lambda|^2$ replaced by
$|{\cal M}^{S,m}_\lambda|^2$. However, the former quantity evaluated in terms
of the latter also involves interference contributions from different
hyperfine amplitudes, which must be disposed of.

In the case of the $(S=1,m=\pm1)$ hyperfine states, this is
straightforward, since one has
\begin{equation}
|{\cal M}^{1,\pm 1}_\lambda|^2=
|{\cal M}_\lambda(\pm\hat{e}_3,\pm\hat{e}_3)|^2,
\end{equation}
where the right-handed orthonormalized basis
$\{\hat{e}_1,\hat{e}_2,\hat{e}_3\}$ with respect to which the
spherical coordinates $(\theta,\varphi)$ are defined, has been introduced.

The situation for the two other hyperfine states $(S=1,m=0)$ and
$(S=0,m=0)$ is more involved however, because of their intertwined
character in terms of the $|\hat{s}_\mu>_\mu|\hat{s}_1>_1$ spin states.
One way to proceed is as follows.

Consider the following specific combinations,
\begin{equation}
X=|{\cal M}_\lambda(\hat{e}_3,-\hat{e}_3)|^2
\,+\,|{\cal M}_\lambda(-\hat{e}_3,\hat{e}_3)|^2=
|{\cal M}^{1,0}_\lambda|^2+|{\cal M}^{0,0}_\lambda|^2,
\end{equation}
\begin{equation}
\begin{array}{r c l}
Y&=&|{\cal M}_\lambda(\hat{e}_1,\hat{e}_1)|^2
-|{\cal M}_\lambda(\hat{e}_1,-\hat{e}_1)|^2
-|{\cal M}_\lambda(-\hat{e}_1,\hat{e}_1)|^2
+|{\cal M}_\lambda(-\hat{e}_1,-\hat{e}_1)|^2\\
&=&|{\cal M}^{1,0}_\lambda|^2-|{\cal M}^{0,0}_\lambda|^2+
2{\rm Re}\left({\cal M}^{1,1}_\lambda{\cal M}^{1,-1}_\lambda\right),
\end{array}
\end{equation}
and
\begin{equation}
\begin{array}{r c l}
Z&=&\frac{\partial^2}{\partial\varphi_\mu\partial\varphi_1}
|{\cal M}_\lambda(\cos\varphi_\mu\hat{e}_1+\sin\varphi_\mu\hat{e}_2,
\cos\varphi_1\hat{e}_1+\sin\varphi_1\hat{e}_2)|^2_{|_{
\varphi_\mu=0,\varphi_1=0}}\\
&=&\frac{1}{4}\left[
|{\cal M}^{1,0}_\lambda|^2-|{\cal M}^{0,0}_\lambda|^2-
2{\rm Re}\left({\cal M}^{1,1}_\lambda{\cal M}^{1,-1}_\lambda\right)\right].
\end{array}
\end{equation}
Then it follows obviously that the remaining two hyperfine state
capture distributions are obtained from
\begin{equation}
|{\cal M}^{1,0}_\lambda|^2=\frac{1}{2}\left[X+\frac{1}{2}(Y+4Z)\right]\ \ ,\ \ 
|{\cal M}^{0,0}_\lambda|^2=\frac{1}{2}\left[X-\frac{1}{2}(Y+4Z)\right]. 
\end{equation}

Applying this procedure to the quantity ${\cal N}_\lambda$ defined
in (\ref{eq:ND}) then leads to the hyperfine state distributions detailed
in Sect.\ref{Sect2.4}. Similarly, the quantity $\vec{\cal D}_\lambda$
defined in (\ref{eq:ND}) would lead to the final nucleus polarization
state associated to capture from each of the muonic hyperfine states,
as given in (\ref{eq:polari}).

\section*{Appendix 3: Left-Right Symmetric Models, 
Leptoquarks and Contact Interactions}
\label{A3}

The purpose of this Appendix is to provide explicit expressions for the
effective coupling coefficients $h^{S,V,T}_{\pm\pm}$
introduced in (\ref{eq:effec}) in terms of
the parameters of specific models or parametrizations for physics
beyond the SM. This allows for model independent bounds 
that could been determined for the $h^{S,V,T}_{\pm\pm}$ coefficients
from some given experiment, to be translated into may be more physically
tangible numbers to be compared with the reach of other experiments,
especially at high energy colliders. Three general classes of such
models beyond the SM are considered here, namely left-right symmetric
models (LRSM), contact interactions
and leptoquarks\cite{Buchmuller,Davinson}.

\vspace{10pt}

\noindent\underline{\sl Left-Right symmetric models}

In the case of LRSM, we refer to the notations, discussion and references in 
Ref.\cite{PDG}. For the process of interest in this paper, in the limit
that Higgs exchange is ignored, which ought to be justified given the
necessarily large masses for such particles in LRSM as well as their
small Yukawa couplings directly proportional to the masses of the
quarks and leptons involved in the process, only $V$ and $A$ couplings
arise in addition to the purely $(V-A)$ character of the SM charged
electroweak interactions. Two massive charged gauges bosons $W^\pm_{1,2}$
appear, with masses $M^W_{1,2}$, which, up to their mixing through an
angle $\zeta$ possibly accompanied by a CP violating phase $\omega$,
are associated to $(V-A)$ and $(V+A)$ interactions characterized by
gauge coupling constants $g_{L,R}$. In the fermionic sector, flavour
mixing is also parametrized by Cabibbo-Kobayashi-Maskawa mixing matrices,
associated to each chirality sector, denoted $V^{L,R}$ and $U^{L,R}$
for the quark and lepton sectors, respectively (indeed in a generic LRSM,
neutrinos are massive and thus flavour-mix with one another as the quarks
do). Let us introduce the following combinations of parameters,
\begin{equation}
\delta=\frac{(M^W_1)^2}{(M^W_2)^2}\ \ ,\ \ 
r=\frac{g_R}{g_L}\ \ ,\ \ 
t=\tan\zeta\ \ ,\ \ 
v_{ud}=\frac{V^R_{ud}}{V^L_{ud}}\ \ ,\ \ 
\rho=\frac{g^2_L}{(M^W_1)^2}\,\cos^2\zeta\,V^{L}_{ud}.
\end{equation}
Given the parametrization (\ref{eq:effec}), one then finds for the
only non vanishing effective coupling coefficients,
\begin{equation}
\frac{g^2}{8M^2}V_{ud}\,h^V_{--}=\rho\left(1+\delta t^2\right)\,U^{L}_{\mu i}
\ \ ,\ \ 
\frac{g^2}{8M^2}V_{ud}\,h^V_{-+}=-\rho rt\left(1-\delta\right)e^{i\omega}
\,U^{L}_{\mu i}v_{ud},
\end{equation}
\begin{equation}
\frac{g^2}{8M^2}V_{ud}\,h^V_{++}=\rho r^2\left(t^2+\delta\right)\,
U^{R}_{\mu i}v_{ud}\ \ ,\ \ 
\frac{g^2}{8M^2}V_{ud}\,h^V_{+-}=-\rho rt\left(1-\delta\right)e^{-i\omega}
\,U^{R}_{\mu i}.
\end{equation}
Here, the indices $i=1,2,3$ and $\mu$ on the neutrino CKM matrix elements
$U^{L,R}_{\mu i}$ stand for the neutrino mass and muon flavour eigenstates,
respectively. 

\vspace{10pt}

\noindent\underline{\sl Contact interactions}

Let us now turn to the parametrization of so-called contact interactions,
which provide a simple-minded model to probe the scale for compositeness
of quarks and leptons. Such interactions are typically represented through
an effective four-fermi coupling of the form\cite{PDG}, say in the case of
vector operators,
\begin{equation}
4\epsilon_{\eta_1\eta_2}\frac{g^2_c}{8\Lambda^2}
\overline{\psi}\gamma_\mu P_{\eta_1}\psi\,
\overline{\psi}\gamma^\mu P_{\eta_2}\psi,
\end{equation}
where $\epsilon_{\eta_1,\eta_2}=\pm1$, $g_c$ is a contact interaction
coupling constant, $\Lambda$ is the associated energy scale, and
$P_{\eta_{1,2}}$ are the chirality projectors already introduced in
(\ref{eq:effec}). In the case of nuclear muon capture, the different
spinor fields appearing in this expression correspond of course to
those of (\ref{eq:effec}), namely the $u$ and $d$ quarks, and the
muon and its associated neutrino. It is conventional\cite{PDG} to
fix the scale $\Lambda$ by setting $g^2_c=4\pi$ (which in the case
of QED would amount to having the fine structure constant set to unity,
$\alpha=1$).

Clearly, such contact interactions may be introduced for any of the
scalar, vector and tensor couplings included in (\ref{eq:effec}),
and for any combination of the fermion chiralities involved.
Thus, one could associate a scale $\Lambda$ say to vector interactions
of $LL$, $VV$, $VA$, etc ... chiralities, in any combination possible,
and similarly for scalar and tensor interactions. Note that in the case
of nuclear muon capture, these contact interactions are all related
to processes which couple the quarks of the first generation to the
leptons of the second.

\vspace{10pt}

\noindent\underline{\sl Leptoquarks}

Finally, let us consider the case of 
leptoquarks (LQ)\cite{Buchmuller,Davinson,PDG}.
These particles come in two varieties, namely either spin 0 or spin 1,
and their name derives from the fact that they couple always to a
quark and a lepton in a single vertex. When one allows for right-handed
neutrinos as well (which was not considered in the original 
discussion\cite{Buchmuller}), there are six different types of scalar
and of vector LQ, characterized by their weak isospin and electric
charge. Their quantum numbers under $SU(3)_c\times SU(2)_L\times U(1)$
are as follows, for scalar LQ,
\begin{equation}
\begin{array}{r l c r l}
S_0: & (3,1,-2/3)\ \ &,&\ \  Q: &(-1/3)\\
\tilde{S}_0: & (3,1,-8/3)\ \ &,&\ \ Q: & (-4/3)\\
\tilde{S}_{0\nu}: & (3,1,4/3)\ \ &,&\ \ Q: & (2/3)\\
S_{1/2}: & (\bar{3},2,-7/3)\ \ &,&\ \ Q: & (-2/3,-5/3)\\
\tilde{S}_{1/2}: & (\bar{3},2,-1/3)\ \ &,&\ \ Q: & (1/3,-2/3)\\
S_1: & (3,3,-2/3)\ \ &,&\ \ Q: & (2/3,-1/3,-4/3)
\end{array}
\end{equation}
and for vector LQ
\begin{equation}
\begin{array}{r l c r l}
V_0: & (\bar{3},1,-4/3)\ \ &,&\ \ Q: & (-2/3)\\
\tilde{V}_0: & (\bar{3},1,-10/3)\ \ &,&\ \ Q: & (-5/3)\\
\tilde{V}_{0\nu}: & (3,1,2/3)\ \ &,&\ \ Q: & (1/3)\\
V_{1/2}: & (3,2,-5/3)\ \ &,&\ \ Q: & (-1/3,-4/3)\\
\tilde{V}_{1/2}: & (3,2,1/3)\ \ &,&\ \ Q: & (2/3,-1/3)\\
V_1: & (\bar{3},3,-4/3)\ \ &,&\ \ Q: & (1/3,-2/3,-5/3)
\end{array}
\end{equation}
where each time the electric charge content of the associated isospin
multiplet is given on the right. Note that the lower index carried by each
of these fields labels its weak isospin value. Moreover,
$\tilde{S}_{0\nu}$ and $\tilde{V}_{0\nu}$ are those LQ related to the
introduction of right-handed neutrinos.

The scalar LQ interactions are then parametrized according to the
Lagrangian density,
\begin{equation}
\begin{array}{r c l}
{\cal L}_S&=&\lambda^L_{S_0}\overline{q^c_L}i\tau_2\ell_LS^\dagger_0\\
 &+&\lambda^R_{S_0}\overline{u^c_R}\mu_RS^\dagger_0+
	\lambda^{R\nu}_{S_0}\overline{d^c_R}\nu_RS^\dagger_0
 +\lambda^R_{\tilde{S}_0}\overline{d^c_R}\mu_R\tilde{S}^\dagger_0+
	\lambda^{R\nu}_{\tilde{S}_{0\nu}}\overline{u^c_R}\nu_R
			\tilde{S}^\dagger_{0\nu}\\
 &+&\lambda^L_{S_{1/2}}\overline{u_R}S^\dagger_{1/2}\ell_L+
	\lambda^R_{S_{1/2}}\overline{q_L}S^\dagger_{1/2}i\tau_2\mu_R
 +\lambda^L_{\tilde{S}_{1/2}}\overline{d_R}\tilde{S}^\dagger_{1/2}\ell_L+
	\lambda^R_{\tilde{S}_{1/2}}\overline{q_L}\tilde{S}^\dagger_{1/2}
			i\tau_2\nu_R\\
 &+&\lambda^L_{S_1}\overline{q^c_L}\vec{S}^\dagger_1\cdot
			i\tau_2\vec{\tau}\ell_L\
	+\ {\rm h.c.}
\end{array}
\end{equation}
and similarly for vector LQ,
\begin{equation}
\begin{array}{r c l}
{\cal L}_V&=&\lambda^L_{V_0}\overline{q_L}\gamma_\mu\ell_L{V^\mu_0}^\dagger\\
 &+&\lambda^R_{V_0}\overline{d_R}\gamma_\mu\mu_R{V^\mu_0}^\dagger+
	\lambda^{R\nu}_{V_0}\overline{u_R}\gamma_\mu\nu_R{V^\mu_0}^\dagger
 +\lambda^R_{\tilde{V}_0}\overline{u_R}\gamma_\mu\mu_R
			\tilde{V}^{\mu\dagger}_0+
	\lambda^{R\nu}_{\tilde{V}_{0\nu}}\overline{d_R}\gamma_\mu\nu_R
			\tilde{V}^{\mu\dagger}_{0\nu}\\
 &+&\lambda^L_{V_{1/2}}\overline{d^c_R}{V^\mu_{1/2}}^\dagger\gamma_\mu\ell_L+
	\lambda^R_{V_{1/2}}\overline{q^c_L}{V^\mu_{1/2}}^\dagger
			\gamma_\mu\mu_R
 +\lambda^L_{\tilde{V}_{1/2}}\overline{u^c_R}\tilde{V}^{\mu\dagger}_{1/2}
			\gamma_\mu\ell_L+
	\lambda^R_{\tilde{V}_{1/2}}\overline{q^c_L}
			\tilde{V}^{\mu\dagger}_{1/2}
			\gamma_\mu\nu_R\\
 &+&\lambda^L_{V_1}\overline{q_L}\vec{V}^{\mu\dagger}_1\cdot
			\gamma_\mu\vec{\tau}\ell_L\
	+\ {\rm h.c.}
\end{array}
\end{equation}
Here, $L$ and $R$ stand for definite chiral components of the spinor fields,
the upper index ``$^c$" refers to the charge conjugate fields, and
$q_L$ and $\ell_L$ stand for the following quark and lepton doublets
\begin{equation}
q_L:\ \left(\begin{array}{c}
		u_L\\
		d_L
		\end{array}\right)\ \ ,\ \ 
\ell_L:\ \left(\begin{array}{c}
		\nu_L\\
		\mu_L
		\end{array}\right).
\end{equation}
Finally, the different $\lambda^{L,R}_{S,V}$ coefficients are complex
constant parameters, the LQ coupling constants. In general, these
coefficients are matrices in generation space, but in the case of
nuclear muon capture, only those LQ couplings between the first 
quark generation and the second lepton generation, including the 
possibility of right-handed neutrinos, are relevant, hence our choice
of notation.

Given these definitions, the induced non vanishing
effective couplings coefficients $h^{S,V,T}_{\pm\pm}$ 
in (\ref{eq:effec}) are expressed as,
\begin{equation}
\frac{g^2}{8M^2}V_{ud}\,\left(h^S_{--}\right)^*=
-\frac{1}{2}\frac{\lambda^{L*}_{V_0}\lambda^{R}_{V_0}}{M^2_{V_0}(-2/3)}
-\frac{1}{2}\frac{\lambda^{L*}_{V_{1/2}}
\lambda^{R}_{V_{1/2}}}{M^2_{V_{1/2}}(-1/3)},
\end{equation}
\begin{equation}
\frac{g^2}{8M^2}V_{ud}\,\left(h^S_{-+}\right)^*=
+\frac{1}{8}\frac{\lambda^{L}_{S_0}\lambda^{R\nu*}_{S_0}}{M^2_{S_0}(1/3)}
+\frac{1}{8}\frac{\lambda^{L}_{\tilde{S}_{1/2}}
\lambda^{R*}_{\tilde{S}_{1/2}}}{M^2_{\tilde{S}_{1/2}}(-2/3)},
\end{equation}
\begin{equation}
\frac{g^2}{8M^2}V_{ud}\,\left(h^S_{+-}\right)^*=
-\frac{1}{8}\frac{\lambda^{L*}_{S_0}\lambda^{R}_{S_0}}{M^2_{S_0}(1/3)}
-\frac{1}{8}\frac{\lambda^{L*}_{S_{1/2}}
\lambda^{R}_{S_{1/2}}}{M^2_{S_{1/2}}(-2/3)},
\end{equation}
\begin{equation}
\frac{g^2}{8M^2}V_{ud}\,\left(h^S_{++}\right)^*=
-\frac{1}{2}\frac{\lambda^{L}_{V_0}\lambda^{R\nu*}_{V_0}}{M^2_{V_0}(-2/3)}
-\frac{1}{2}\frac{\lambda^L_{\tilde{V}_{1/2}}
\lambda^{R*}_{\tilde{V}_{1/2}}}{M^2_{\tilde{V}_{1/2}}(-1/3)},
\end{equation}
\begin{equation}
\frac{g^2}{8M^2}V_{ud}\,\left(h^V_{--}\right)^*=
\frac{g^2_L}{8M^2_W}V^{\rm SM}_{ud}
+\frac{1}{8}\frac{|\lambda^{L}_{S_0}|^2}{M^2_{S_0}(1/3)}
-\frac{1}{8}\frac{|\lambda^{L}_{S_1}|^2}{M^2_{S_1}(-1/3)}
+\frac{1}{4}\frac{|\lambda^{L}_{V_0}|^2}{M^2_{V_0}(-2/3)}
-\frac{1}{4}\frac{|\lambda^{L}_{V_1}|^2}{M^2_{V_1}(-2/3)},
\end{equation}
\begin{equation}
\frac{g^2}{8M^2}V_{ud}\,\left(h^V_{++}\right)^*=
-\frac{1}{8}\frac{\lambda^{R}_{S_0}\lambda^{R\nu*}_{S_0}}{M^2_{S_0}(1/3)}
+\frac{1}{4}\frac{\lambda^{R}_{V_0}\lambda^{R\nu*}_{V_0}}{M^2_{V_0}(-2/3)},
\end{equation}
\begin{equation}
\frac{g^2}{8M^2}V_{ud}\,\frac{1}{2}\left(h^T_{-+}\right)^*=
-\frac{1}{32}\frac{\lambda^{L}_{S_0}\lambda^{R\nu*}_{S_0}}{M^2_{S_0}(1/3)}
+\frac{1}{32}\frac{\lambda^{L}_{\tilde{S}_{1/2}}
\lambda^{R*}_{\tilde{S}_{1/2}}}{M^2_{\tilde{S}_{1/2}}(-2/3)},
\end{equation}
\begin{equation}
\frac{g^2}{8M^2}V_{ud}\,\frac{1}{2}\left(h^T_{+-}\right)^*=
+\frac{1}{32}\frac{\lambda^{L*}_{S_0}\lambda^{R}_{S_0}}{M^2_{S_0}(1/3)}
-\frac{1}{32}\frac{\lambda^{L*}_{S_{1/2}}\lambda^{R}_{S_{1/2}}}
{M^2_{S_{1/2}}(-2/3)}.
\end{equation}
In these expressions, which of the LQ isospin component contributes
to a each coefficient is indicated by giving its electric charge
in the parenthesis following the mass value in the mass contributions
$1/M^2_{S,V}$. Note that the $\tilde{S}_{0\nu}$ and $\tilde{V}_{0\nu}$
LQ do not contribute to these coefficients.

\clearpage

\begin{table}
\begin{center}
\begin{tabular}{|c||l|l|l|l|l|l|}
\hline
 & & & & & & \\
 & $\sigma(\lambda^{\rm S};F_X)$ &
$\sigma(\lambda^{\rm T};F_X)$ &
$\sigma(\lambda^{\rm stat};F_X)$ &
$\sigma(A_\Delta;F_X)$ & 
$\sigma(A_v;F_X)$ & 
$\sigma(A_t;F_X)$ \\
 & & & & & & \\
\hline
$F_V$ & $-0.726$ & $+0.790$ & $+0.301$ & $-4.572$ & $+0.798$ & $+0.0616$ \\
\hline
$F_M$ & $+0.420$ & $+0.233$ & $+0.293$ & $+0.561$ & $-0.287$ & $+0.254$ \\
\hline
$F_A$ & $+2.621$ & $+0.997$ & $+1.521$ & $+4.894$ & $-0.134$ & $-1.063$ \\
\hline
$F_P$ & $-0.314$ & $-0.0210$ & $-0.116$ & $-0.884$ & $-0.377$ & $+0.747$ \\
\hline
\hline
$F_S$ & $-0.0155$ & $+0.0178$ & $+0.0071$ & $-0.101$ & $+0.0173$ 
	& $+0.0017$ \\
\hline
$F_T$ & $-0.0157$ & $-0.00112$ & $-0.0058$ & $-0.044$ & $-0.0185$ & $+0.0368$ \\
\hline
\end{tabular}
\caption[]{Sensitivities of the different observables 
${\cal O}=\lambda^{\rm S,T,stat},A_\Delta,A_v,A_t$ for muon capture
on $^3$He, with respect to 
the nuclear form factors associated to the vector and axial quark 
operators. See (\ref{eq:Sen}) for the definition of 
$\sigma({\cal O};F_X)$.}
\label{Table1}
\end{center}
\end{table}

\begin{table}
\begin{center}
\begin{tabular}{|c||l|l|l|l|l|l|}
\hline
 & & & & & & \\
 & $\sigma(\lambda^{\rm S};h_X)$ &
$\sigma(\lambda^{\rm T};h_X)$ &
$\sigma(\lambda^{\rm stat};h_X)$ &
$\sigma(A_\Delta;h_X)$ & 
$\sigma(A_v;h_X)$ & 
$\sigma(A_t;h_X)$ \\
 & & & & & & \\
\hline
$h^S_{+-}$ & $-0.840$ & $+0.946$ & $+0.370$ & $-5.38$ & $+0.901$ & $+0.126$ \\
\hline
$h^S_{++}$ & $-0.810$ & $+0.948$ & $+0.381$ & $-5.30$ & $+0.937$ & $+0.0544$ \\
\hline
$h^S_+$ & $-0.825$ & $+0.947$ & $+0.375$ & $-5.34$ & $+0.919$ & $+0.0902$ \\
\hline
$h^P_+$ & $-0.0151$ & $-0.00093$ & $-0.00563$ & $-0.0426$ & $-0.0180$ &
$+0.0361$ \\
\hline
$\frac{1}{2}h^T_{+-}$ & 
$+10.1$ & $+3.78$ & $+5.82$ & $+19.00$ & $-0.733$ & $-3.83$ \\
\hline
$h^V_{--}$ & $+2.00$ & $+2.00$ & $+2.00$ & $+0.00$ & $+0.00$ & 
$+0.00$ \\
\hline
$h^V_{-+}$ & $-2.61$ & $+0.0478$ & $-0.810$ & $-8.022$ & $+1.022$ &
$+0.632$ \\
\hline
\end{tabular}
\caption[]{Sensitivities of the different observables 
${\cal O}=\lambda^{\rm S,T,stat},A_\Delta,A_v,A_t$ for muon capture
on $^3$He, with respect to 
the effective coupling coefficients $h^{S,V,T}_{\pm\pm}$.
Sensitivities for those couplings which do not appear in this table
are identically vanishing. The scalar and pseudoscalar combinations are
defined by $h^S_+=h^S_{+-}+h^S_{++}$ and $h^P_+=h^S_{+-}-h^S_{++}$, while
it is understood that all form factors $G_S$, $G_P$ and $G_T$ are set
to unity to obtain the numbers in this table.
See (\ref{eq:Sen}) for the definition of 
$\sigma({\cal O};h_X)$.}
\label{Table2}
\end{center}
\end{table}

\begin{table}
\begin{center}
\begin{tabular}{|c||l|l|l|l|l|l|}
\hline
 & & & & & & \\
 & $\sigma(\lambda^{\rm S};g_X)$ &
$\sigma(\lambda^{\rm T};g_X)$ &
$\sigma(\lambda^{\rm stat};g_X)$ &
$\sigma(A_\Delta;g_X)$ & 
$\sigma(A_v;g_X)$ & 
$\sigma(A_t;g_X)$ \\
 & & & & & & \\
\hline
$g_V$ & $+0.466$ & $-1.129$ & $+0.386$ & $+0.108$ & $+13.16$ & $-2.39$ \\
\hline
$g_M$ & $+0.151$ & $+0.680$ & $+0.177$ & $-0.0357$ & $-0.317$ & $+0.551$ \\
\hline
$g_A$ & $+1.567$ & $+1.440$ & $+1.561$ & $+0.00856$ & $-18.85$ & $+0.991$ \\
\hline
$g_P$ & $-0.184$ & $+1.008$ & $-0.125$ & $-0.0804$ & $+6.01$ & $+0.844$ \\
\hline
\hline
$g_S$ & $+0.0232$ & $-0.0718$ & $+0.0185$ & $-0.00641$ & $+0.724$ 
	& $-0.139$ \\
\hline
$g_T$ & $+0.0238$ & $-0.125$ & $+0.0164$ & $+0.01003$ & $-0.759$ & $-0.105$ \\
\hline
\end{tabular}
\caption[]{Sensitivities of the different observables 
${\cal O}=\lambda^{\rm S,T,stat},A_\Delta,A_v,A_t$ for muon capture
on the proton,  with respect to 
the nuclear form factors associated to the vector and axial quark 
operators. See (\ref{eq:Sen}) for the definition of 
$\sigma({\cal O};F_X)$.}
\label{Table3}
\end{center}
\end{table}

\begin{table}
\begin{center}
\begin{tabular}{|c||l|l|l|l|l|l|}
\hline
 & & & & & & \\
 & $\sigma(\lambda^{\rm S};h_X)$ &
$\sigma(\lambda^{\rm T};h_X)$ &
$\sigma(\lambda^{\rm stat};h_X)$ &
$\sigma(A_\Delta;h_X)$ & 
$\sigma(A_v;h_X)$ & 
$\sigma(A_t;h_X)$ \\
 & & & & & & \\
\hline
$h^S_{+-}$ & $+0.434$ & $-1.394$ & $+0.343$ & $+0.123$ & $+12.16$ & $-2.562$ \\
\hline
$h^S_{++}$ & $+0.391$ & $-1.157$ & $+0.314$ & $+0.104$ & $+13.57$ & $-2.363$ \\
\hline
$h^S_+$ & $+0.412$ & $-1.274$ & $+0.328$ & $+0.114$ & $+12.86$ & $-2.461$ \\
\hline
$h^P_+$ & $+0.0217$ & $-0.118$ & $+0.0147$ & $+0.00945$ & $-0.710$ &
$-0.0986$ \\
\hline
$\frac{1}{2}h^T_{+-}$ & 
$-5.21$ & $-5.58$ & $-5.23$ & $+0.0251$ & $+55.21$ & $-3.65$ \\
\hline
$h^V_{--}$ & $+2.00$ & $+2.00$ & $+2.00$ & $+0.00$ & $+0.00$ & 
$+0.00$ \\
\hline
$h^V_{-+}$ & $-0.767$ & $-2.90$ & $-0.872$ & $+0.144$ & $+25.68$ &
$-3.669$ \\
\hline
\end{tabular}
\caption[]{Sensitivities of the different observables 
${\cal O}=\lambda^{\rm S,T,stat},A_\Delta,A_v,A_t$ for muon capture
on the proton,  with respect to 
the effective coupling coefficients $h^{S,V,T}_{\pm\pm}$.
Sensitivities for those couplings which do not appear in this table
are identically vanishing. The scalar and pseudoscalar combinations are
defined by $h^S_+=h^S_{+-}+h^S_{++}$ and $h^P_+=h^S_{+-}-h^S_{++}$, while
it is understood that all form factors $G_S$, $G_P$ and $G_T$ are set
to unity to obtain the numbers in this table.
See (\ref{eq:Sen}) for the definition of 
$\sigma({\cal O};h_X)$.}
\label{Table4}
\end{center}
\end{table}

\end{document}